\documentclass[11pt,a4paper]{article}

\usepackage{jheppub}

\usepackage[utf8]{inputenc}

\usepackage{graphicx}
\usepackage{amsmath,amsthm,amssymb}

\newcommand{\id}{\mathbb{1}}
\newcommand{\ie}{\emph{i.e.}, }
\newcommand{\tr}{\operatorname{tr}}
\newcommand{\ii}{\operatorname{i}}
\usepackage{braket}
\usepackage[bb=boondox]{mathalfa}
\usepackage{tikz}
\usetikzlibrary{decorations.markings,arrows,decorations.pathreplacing,calc,intersections,through,backgrounds}
\newcommand\manifold[3][]{
	\draw[every to/.style={out=-20,in=160,relative},#1] (#2) 
	to ($(#2 -| #3)!0.2!(#2 |- #3)$)
	to (#3)
	to ($(#2 -| #3)!0.8!(#2 |- #3)$)
	to cycle;
}

\tikzset{->-/.style={decoration={markings,mark=at position #1 with {\arrow{>}}},postaction={decorate}}}
\tikzset{
	partial ellipse/.style args={#1:#2:#3}{
		insert path={+ (#1:#3) arc (#1:#2:#3)}
	}
}

\makeatletter
\newsavebox{\@brx}
\newcommand{\Mod}[1]{\ (\mathrm{mod}\ #1)}
\newcommand{\llangle}[1][]{\savebox{\@brx}{\(\m@th{#1\langle}\)}%
  \mathopen{\copy\@brx\mkern2mu\kern-0.9\wd\@brx\usebox{\@brx}}}
\newcommand{\rrangle}[1][]{\savebox{\@brx}{\(\m@th{#1\rangle}\)}%
  \mathclose{\copy\@brx\mkern2mu\kern-0.9\wd\@brx\usebox{\@brx}}}
\makeatother

\makeatletter
\def\@fpheader{\relax}
\makeatother

\newcommand{\diff}[1]{\frac{\partial}{\partial #1}}
\newcommand{\diffn}[2]{\frac{\partial^{#2}}{\partial #1^{#2}}}
\newcommand{\dJ}{\diff{J}}
\newcommand{\dJn}[1]{\diffn{J}{#1}}

\title{Towards Spacetime Entanglement Entropy for Interacting Theories}

\author[a]{Yangang Chen,}
\author[b]{Lucas Hackl,}
\author[c,d]{Ravi Kunjwal,}
\author[e,f]{Heidar Moradi,}
\author[g,h]{Yasaman~K.~Yazdi,}
\author[i]{and Miguel~Zilh\~{a}o}

\affiliation[a]{Department  of  Applied  Mathematics,  University  of  Waterloo,\\200  University  Avenue  West, Waterloo, ON, N2L 3G1, Canada}
\affiliation[b]{QMATH,  Department  of  Mathematical  Sciences,  University  of  Copenhagen,
Universitetsparken  5,  2100  Copenhagen,  Denmark}
\affiliation[c]{Perimeter Institute for Theoretical Physics,\\31 Caroline St. N, Waterloo, Ontario, N2L 2Y5, Canada}
\affiliation[d]{Centre for Quantum Information and Communication, Ecole polytechnique de Bruxelles,\\CP 165, Universit\'e libre de Bruxelles, 1050 Brussels, Belgium}
\affiliation[e]{Department of Applied Mathematics and Theoretical Physics, Centre for Mathematical Sciences,\\Wilberforce Road, Cambridge CB3 0WA, UK}
\affiliation[f]{The Cavendish Laboratory, Department of Physics,\\19 J J Thomson Avenue, Cambridge CB3 0HE, UK}
\affiliation[g]{Department of Physics, 4-181 CCIS, University of Alberta,\\ Edmonton AB, T6G 2E1, Canada}
\affiliation[h]{Theoretical  Physics  Group, Blackett  Laboratory,  Imperial  College London, SW7 2AZ, UK}
\affiliation[i]{CENTRA,  Departamento  de  F\'{i}sica,  Instituto  Superior  T\'{e}cnico,
Universidade  de  Lisboa,\\Avenida  Rovisco  Pais  1,  1049  Lisboa,  Portugal}

\emailAdd{y493chen@uwaterloo.ca}
\emailAdd{lucas.hackl@math.ku.dk}
\emailAdd{rkunjwal@ulb.ac.be}
\emailAdd{hm598@cam.ac.uk}
\emailAdd{ykouchek@imperial.ac.uk}
\emailAdd{miguel.zilhao.nogueira@tecnico.ulisboa.pt}

\abstract{Entanglement entropy of quantum fields in gravitational settings is a topic of growing importance. 
This entropy of entanglement is conventionally computed relative to Cauchy hypersurfaces where it is possible via a partial tracing to associate a reduced density matrix to the spacelike region of interest. In recent years Sorkin has proposed an alternative, manifestly covariant, formulation of entropy in terms of the spacetime two-point correlation function. This formulation, developed for a Gaussian scalar field theory, is explicitly spacetime in nature and evades some of the possible non-covariance issues faced by the conventional formulation. In this paper we take the first steps towards extending Sorkin's entropy to non-Gaussian theories where Wick's theorem no longer holds and one would expect higher correlators to contribute. We  consider quartic perturbations away from the Gaussian case and find that to first order in perturbation theory, the entropy formula derived by Sorkin continues to hold but with the two-point correlators replaced by their perturbation-corrected counterparts. We then show that our results continue to hold for arbitrary perturbations (of both bosonic and fermionic theories). This is a non-trivial and, to our knowledge, novel result. Furthermore we also derive closed-form formulas of the entanglement entropy for arbitrary perturbations at first and second order.
Our work also suggests avenues for further extensions to   generic  interacting theories.}

\begin{document}
\maketitle

%\today

%\tableofcontents

\section{Introduction}
Around 1952, Rudolf Peierls found himself ``looking for a new expression of some of the basic rules of quantum mechanics, namely the formulation of commutation laws of relativistic field theory''~\cite{peierlsvol2}. There already existed consistent formulations of relativistic quantum field theory, developed decades earlier by Heisenberg and Pauli \cite{Heisenberg1929,Heisenberg1930}, as well as more recent interaction picture frameworks developed by Schwinger \cite{Schwinger}, Tomonaga \cite{Tomonaga} and others. Despite this, Peierls was worried about the Lorentz invariance of these formulations. He wanted to find an alternative approach in which the Lorentz invariance of the formulation is manifestly evident. An unsatisfactory feature of the existing approaches was that a Hamiltonian and its associated canonical variables were needed, thus tying the formalisms to a choice of time coordinate or frame.

Around this time, Peierls succeeded in deriving the covariant formulation of a relativistic quantum field theory that he had in mind. His formulation, explained in \cite{peierls}, determines a general rule for the  commutator at any pair of spacetime points. This formulation uses Green functions of the theory and makes no reference to a Hamiltonian. More precisely, the spacetime commutator  (also known as the Peierls bracket or Pauli Jordan function) is the difference between the retarded and advanced Green functions: $\Delta=G_R-G_A$. Peierls also held much correspondence about his work with other prominent physicists of his time. When his %\cite{peierls} 
paper came out, many were interested in it but also expressed skepticism.\footnote{Notably, Pauli who was very much interested in this work was skeptical about it and stated that he thought ``it looks like a cemetery for the lorentzinvariant formfactor theory in its present form"!\cite{peierlsvol2}} Nonetheless, the Peierls bracket has stood the test of time and is an important framework for quantum field theory that shows that its time-dependence can be captured in a Lorentz invariant manner. It has also enabled important advances in algebraic quantum field theory \cite{kasia1, feden1}, quantum field theory in curved spacetime \cite{feden2, Hollands, Aslanbeigi}, and quantum gravity \cite{SJ1, Histories, InteractingCS}. We will return to the Peierls bracket later in the main text. For now we turn to an important topic in the current century and to the work of another physicist adamant about manifest Lorentz invariance.

Entanglement entropy has become an increasingly important and useful topic in the past few decades. In this work, we will mainly be interested in it in the context of quantum field theory. The physical content of a quantum field theory is often expressed in terms of the set of its $n$-point correlations. Under certain circumstances, either intentionally (often the case in condensed matter systems) or unintentionally (often the case in gravitationally interesting systems such as black hole and cosmological spacetimes), one may not have access to the quantum field in the entire  spacetime in which it lives. As a result of this limited access, we lose the information that was contained in the $n$-point correlations involving points both in the accessible and inaccessible regions. This is where entanglement entropy comes into play. Entanglement entropy is a measure of this loss of information.  

The prototypical and perhaps most interesting background spacetime in which we would not have access to all the correlation information of a quantum field is that of a black hole. The event horizon of the black hole is a boundary that separates the degrees of freedom we have access to and those that we do not. In fact, the birth of the concept of the entanglement entropy of a quantum field occurred exactly as it was studied by Rafael Sorkin in this context, back in 1983 \cite{sorkin1983}. In this first work on the topic, a scalar field in a black hole spacetime was considered. The field data on a Cauchy surface was divided into the interior and exterior (of the black hole) components, and the entanglement entropy 
\begin{equation}\label{vNe}
S=\tr \rho_{\rm ext} \log \rho_{\rm ext}^{-1}    
\end{equation}
was computed.  $\rho_{\rm ext}$ is what remains after the interior data is traced out. It was found that the resulting entropy is proportional to the spatial area of the event horizon in units of the cutoff, i.e.\ $A/\ell^2$ in $4$d. Without a cutoff the entropy is infinite. Because of this scaling property, entanglement entropy was proposed by Sorkin to be a candidate for the microscopic origin of black hole entropy, which also scales like the area of the event horizon.

The question of the microscopic origin of black hole entropy has been the subject of intense research ever since the semiclassical calculations of  Bekenstein and Hawking~\cite{tj, Solodukhin, Emparan}. As mentioned, these calculations showed that the black hole entropy is proportional to the spatial area of the event horizon. An understanding of the microscopic degrees of freedom giving rise to this entropy\footnote{For example a statistical mechanical understanding of this entropy of the form $S\sim\ln(\text{number of microstates})$.} has, however, been elusive. Since Sorkin's proposal, entanglement entropy has been taken seriously as a/the source of the entropy. Numerous studies have already been made on
the connection between entanglement entropy and black hole entropy \cite{bombelli, tj,Solodukhin, Emparan}, but there is not yet a definitive answer regarding whether or not it is the fundamental description of black hole entropy. 

A difficulty in this regard is that the formula 
\begin{equation}\label{vNe2}
S=\tr \rho \log \rho^{-1}    
\end{equation}
is used by defining the density matrix relative to a spatial Cauchy hypersurface $\Sigma$. Similarly, the cutoff that renders the entropy finite and quantifies it is defined relative to this hypersurface. Harking back to the worries of Peierls in 1952, such a construction of entanglement entropy lacks Lorentz invariance. Sorkin was also worried about this and in 2012 \cite{rssee} he derived a covariant definition of spacetime entropy (including but more general than entanglement entropy), using none other than the Peierls bracket. The other major ingredient in this definition is the spacetime correlation function.\footnote{See \cite{sorkingreen} for a prescription for obtaining the correlation function from the Peierls bracket as well.} This definition is reviewed in Section \ref{gaussianSee}.

This brings us to the topic of the current paper. The definition given in \cite{rssee} is limited to the Gaussian theory. A question that arises is whether  a covariant spacetime definition of entropy can also be found for non-Gaussian and interacting theories. At first sight it might seem like a difficult task. In the Gaussian theory, one has the ease of working with only the  two-point function. For a non-Gaussian and/or interacting field theory, one may have to consider the higher $n$-point functions as well. In this case, there are two possibilities for us to succeed in the generalization we seek: i) we find a natural generalization of the formula in \cite{rssee} (possibly a(n) (in)finite set of formulas) that includes the contributions from all $n$-point functions, or ii) we find that not all $n$-point functions contribute to the entropy. 

In the present work, we find that up to first order in perturbation theory possibility (ii) holds: two-point functions suffice to capture the entropy. Beyond first order, we are faced with possibility (i). In particular we show explicitly how all higher-order correlators are needed to fully capture the entropy at second order. We begin by considering a generic non-Gaussian scalar theory with quartic perturbations to the quadratic Gaussian density matrix and we arrive at a formula for the entropy that is essentially the same as the one in \cite{rssee} up to first order. The difference in our case is that the  perturbation-corrected two-point correlation function enters the formula rather than the unperturbed one from the Gaussian theory. We then prove that the same formula captures the first order contribution to the entanglement entropy for \emph{any}  perturbation away from a (bosonic or fermionic) Gaussian theory. This finding may also point towards a deeper understanding of the information content of the (entanglement) entropy of a quantum field. At second order, we find that a part of the entropy is still captured by the same formula as the previous two orders, but the full second order contribution is not captured by it. We show that generically the full second order correction contains contributions of all higher-order correlation functions. Furthermore, we derive closed-form formulas for the entropy for arbitrary perturbations at first and second order.
These findings can facilitate several extensions of the current work to more complicated  interacting theories, in particular the extension of the formula in \cite{rssee}.

\section{General Quantum Field Theory for a Real Scalar Theory}
In this section we briefly discuss aspects of a general quantum field theory with a real scalar field. Since we are interested in a spacetime formulation we will work in the Heisenberg picture, loosely using the language of its axiomatizations such as the Haag-Kastler \cite{Haag, Haag2} and Wightman axioms \cite{Wightman, Wightman2}. We will not attempt to make this presentation rigorous in any way. For further details, see \cite{HollandsSanders, Fewster:2019ixc, WittenEntanglement, AQFTcurvedspacetime,Fredenhagen:2012sb}.

Given a spacetime $(\mathcal M,g)$ with manifold $\mathcal M$ and metric $g$, for any region $\mathcal R\subset\mathcal M$ we can associate a unital $\star$-algebra of observables called $\mathcal A_{\mathcal R}$. These algebras must satisfy certain properties. For example for any subregion $\mathcal R_1\subset\mathcal R_2$ the corresponding algebras are nested $\mathcal A_{\mathcal R_1}\subset\mathcal A_{\mathcal R_2}$. In order to ensure causality, the algebras for any causally disjoint (with respect to the metric $g$) regions $\mathcal R_1$ and $\mathcal R_2$ must commute, i.e.
\begin{equation}
    [\mathcal O_1, \mathcal O_2] = 0, \qquad \forall \mathcal O_1\in A_{\mathcal R_1}, \quad \text{and}\quad \forall\mathcal O_2\in A_{\mathcal R_2}.
\end{equation}
It is also required that if $\mathcal R_1$ contains a Cauchy surface $\Sigma$ of $\mathcal R_2$, then
\begin{equation}
    \mathcal A_{\mathcal R_1} = \mathcal A_{\mathcal R_2}.
\end{equation}
This requirement is essentially about the existence of dynamics. The logic is that observables outside the Cauchy surface $\Sigma$ are determined by dynamical time evolution of the theory. In other words, operators with support in $\mathcal R_2$ outside $\mathcal R_1$ can be constructed as (fairly complicated) functions of operators in $\mathcal A_{\mathcal R_1}$ or just $A_{\Sigma}$.
Other requirements include some form of automorphism under Poincar\'e transformations in Minkowski spacetime, or a compatibility axiom in more general spacetimes \cite{AQFTcurvedspacetime}. The set of assignments $\mathcal R\rightarrow \mathcal A_{\mathcal R}$ is sometimes called a net of local algebras. For a nested family of regions that satisfy $\cup_n\mathcal R_n = \mathcal M$, the net of local algebras can be used to construct the full spacetime algebra by $\cup_n\mathcal A_{\mathcal R_n} = \mathcal A_{\mathcal M}$.\footnote{Or by its closure under some appropriate topology.}

In addition to the algebra of observables, we also need a positive functional $\langle\cdot\rangle:\mathcal A\rightarrow\mathbb C$, such that $\langle 1\rangle = 1$, usually called a functional state. In more conventional language this functional assigns expectation values to any given observable in $\mathcal A$, and can therefore be used to the construct correlation functions of the QFT.

It is useful to make the above construction more explicit.\footnote{This abstract construction is often preferred as the Stone-von Neumann theorem fails in QFT and there can be inequivalent unitary representations of the algebra. As we are interested in entanglement of states, it is nonetheless useful to work with explicit Hilbert spaces and ignore potential subtleties.} We will from now on think of the algebras $\mathcal A_{\mathcal R}$ as being generated by a Hermitian scalar field $\phi(x)$ and the expectation values from a given  vacuum state $|0\rangle$. We can then generate the Hilbert space (up to other superselection sectors) by the span of states of the form\footnote{Note that the operator $\phi(x)$ is a distribution-valued operator and there are problems with multiplying these in order to generate $\mathcal A_{\mathcal R}$. In a more careful approach we could, for any bounded test function $f$, define the smeared operators
$\phi(f) = \int_{\mathcal M} d^dx f(x)\phi(x)$ and work with these. In the following we will be less precise and work with products of $\phi(x)$.
}
\begin{equation}
    |\psi\rangle= \phi(x_1)\dots\phi(x_n) |0\rangle,\qquad x_i\in\mathcal R.
\end{equation}
In order to construct this Hilbert space, we do not need the full spacetime algebra $\mathcal A_{\mathcal M}$.
Thinking classically, we would expect that operators on a Cauchy surface $\Sigma$ would be enough to generate the full Hilbert space.
But surprisingly according to the Reeh-Schleider theorem \cite{schlieder1965, Reeh1961}, in quantum field theory much less is needed; any open set $\mathcal R$ is sufficient to generate a dense subspace of $\mathcal H$.

\subsection{Spacetime Entanglement Entropy}
We now discuss what the entanglement entropy of a state $\rho$ is relative to a spacetime region $\mathcal R$. In standard quantum mechanics we have a Hilbert space $\mathcal H$ associated to a Cauchy surface $\Sigma$, where the algebra $\mathcal A_{\Sigma}$ acts irreducibly on $\mathcal H$, and a ``vacuum" state $\rho$. Since the full spacetime $\mathcal M$ is the domain of dependence of $\Sigma$, we have that $\mathcal A_{\mathcal M} = \mathcal A_{\Sigma}$. Therefore $\mathcal A_{\mathcal M}$ acts irreducibly on $\mathcal H$ and we can think of $\rho$ as the global state.

\begin{figure}
    \centering
    \includegraphics[width=0.70\textwidth]{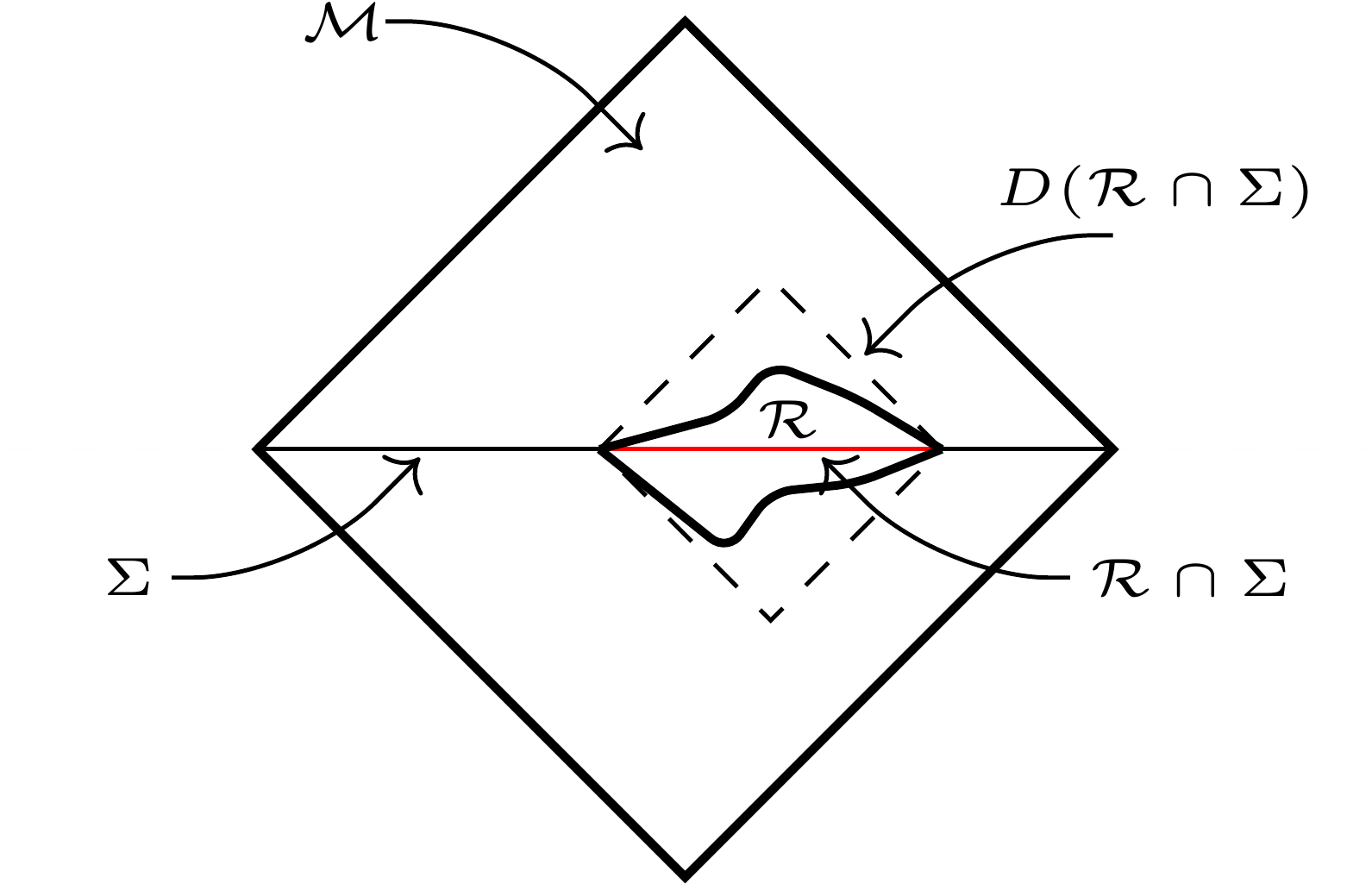}
    \caption{This figure illustrates how the spacetime entropy $S(\mathcal R)$ is related to the conventional Cauchy surface entanglement entropy $S(\Sigma')$, where $\Sigma'\subset\Sigma$. %
    Consider the region $\mathcal R\subset\mathcal M$ such that $\Sigma_{\mathcal R}\equiv \mathcal R\cap\Sigma$ is a Cauchy surface of $\mathcal R$ (or alternatively, $\mathcal R$ is contained within the domain of dependence $D(\mathcal R\cap\Sigma)$). For such regions the entropy $S(\mathcal R)$ corresponds to the entropy  $S(\Sigma'=\Sigma_{\mathcal R})$, which captures the bipartite entanglement entropy between $\Sigma_{\mathcal R}$ and its complement in $\Sigma$.}
    \label{fig:SpacetimeDiagramEntanglement}
\end{figure}

Now consider the subregion $\mathcal R\subset\mathcal M$. In general, we cannot expect the subalgebra $\mathcal A_{\mathcal R}$ to act irreducibly on the Hilbert space $\mathcal H$. However, imagine that we can find another Hilbert space $\mathcal H_{\mathcal R}$ where it does act irreducibly and a density matrix $\rho_{\mathcal R}$ in $\mathcal H_{\mathcal R}$ such that
\begin{equation}
    \left\langle\mathcal O\right\rangle = \tr_{\mathcal H}\left(\rho\mathcal O\right) = \tr_{\mathcal H_{\mathcal R}}\left(\rho_{\mathcal R}\mathcal O\right),\qquad \forall\mathcal O\in\mathcal A_{\mathcal R}.
\end{equation}
In such a case we can define the entropy of $\rho$ relative to the spacetime region $\mathcal R$ as \cite{rssee}
\begin{equation}
    S(\mathcal R) = -\tr_{\mathcal H_{\mathcal R}}\left(\rho_{\mathcal R}\log\rho_{\mathcal R}\right).
\end{equation}
We can think of $S(\mathcal R)$ as a generalized entropy relative to any spacetime region $\mathcal R$. In special cases this can be interpreted as the conventional entanglement entropy; if there exists a Cauchy surface $\Sigma$ of the spacetime $\mathcal M$, such that $\mathcal R\cap\Sigma$ is a Cauchy surface for $\mathcal R$, then $S(\mathcal{R})$ corresponds to the standard bipartite entanglement entropy (see Figure \ref{fig:SpacetimeDiagramEntanglement}). Even in such cases, $S(\mathcal R)$ has the advantage over $S(\mathcal R\cap\Sigma)$ of allowing for a covariant spacetime cutoff. The cutoff plays a central role in the definition of the entropy in field theory; without it the entropy would be  infinite.

\subsection{Quantum Peierls Brackets}
Let $\phi(x)$ be a real scalar field, $\mathcal A$ an algebra generated by $\phi(x)$, $\mathcal H$ a Hilbert space on which $\mathcal A$ acts irreducibly, and $\rho$ a density matrix on this Hilbert space.
From the associative product on $\mathcal A$ we can define a general commutator of the generators $\phi(x)$ such that $[\phi(x),\phi(y)]\in\mathcal A$. In particular (using the notation $\phi^i$ for $\phi(x)$) we generally have something of the form
\begin{gather}\label{eq:generalCommutator}
    \begin{aligned}
        \left[\phi^i,\phi^j\right] &= \sum_{n=0}^\infty i\Delta^{ij, \mathcal O_1\dots \mathcal O_n}_{(n)\;i_1\dots i_n}\sum_{\mathcal O_1,\dots, \mathcal O_n\in\mathcal L}\mathcal O_1^{i_1}\dots \mathcal O_n^{i_n}\\
        &= i\Delta^{ij}_{(0)} + i\sum_{\mathcal O\in\mathcal L}\Delta^{ij, \mathcal O}_{(1)\;k}\mathcal O^k + i\sum_{\mathcal O_1,\mathcal O_2\in\mathcal L}\Delta^{ij}_{(2)\;kl} \mathcal O_1^k \mathcal O_2^l + \dots,
    \end{aligned}
\end{gather}
where the indices $i$ in $\mathcal O^i$ stand for various indices $i = (x, \mu_1, \mu_2, \dots)$, $\mathcal L\subset\mathcal A$ is the set of independent local operators generated from $\phi(x)$, and $\Delta^{ij, \mathcal O_1\dots\mathcal O_n}_{(n)\;i_1\dots i_n}$ are c-numbers.
Local operators can in principle be constructed as sums of products of derivatives of $\phi$\footnote{Note that products of operators at a single point are often singular. In reality, these products must be defined in a more careful manner. In $1+1$ dimensional CFTs this is done using point-split regularization and a generalized normal-ordering prescription. More generally, this can be remedied by smearing.}
\begin{equation}
    \mathcal L = \left\{\phi(x), \partial_\mu\phi(x), \phi^2(x), \mathcal \phi\partial_\mu\phi(x), \partial^\mu\phi\partial_\mu\phi(x), \dots\right\}.
\end{equation}
Note that not all of these operators are scalars; they can be higher-rank tensors as well.
The c-numbers $\Delta^{ij, \mathcal O_1\dots\mathcal O_n}_{(n)\;i_1\dots i_n}$ must naturally satisfy a set of constraints, such as causality and the Jacobi identity.

The spacetime commutators \eqref{eq:generalCommutator} can be thought of as the commutators of operators in the Heisenberg picture, or as quantum Peierls brackets.

Let us define the Wightman correlation tensors for any positive integer $n\in\mathbb N$ as
\begin{equation}
    W_{(n)}^{i_1\dots i_n} = \left\langle\phi^{i_1}\dots\phi^{i_n}\right\rangle.
\end{equation}
In general we expect these to determine the theory fully.
Due to hermiticity of $\phi^i$, we have the conditions
\begin{equation}\label{eq:correlationTensors}
    \overline W_{(n)}^{i_1\dots i_n} = W_{(n)}^{i_n\dots i_1},
\end{equation}
where $\overline{\phantom{W}}$ is complex conjugation. For the two-point correlation matrix this condition is the standard hermiticiy condition $W_{(2)}^\dagger = W_{(2)}$.
Taking the expectation value of the commutator we find that
\begin{gather}
    \begin{aligned}
        W^{ij}_{(2)}-W^{ji}_{(2)} &= i\Omega^{ij}\\
                                  &= i\Delta^{ij}_{(0)} + i\sum_{\mathcal O\in\mathcal L}\Delta^{ij, \mathcal O}_{(1)\;k}\langle\mathcal O^k\rangle + \cdots,
    \end{aligned}
\end{gather}
where $i\Omega^{ij}$ is the expectation value of the right hand side of \eqref{eq:generalCommutator}. For a free theory we have that commutators of the Heisenberg operators are c-numbers \cite{Baumann, Greenberg} (see also Appendix \ref{appendix:c_numberStuff}) and thus
\begin{equation}
    W^{ij}_{(2)}-W^{ji}_{(2)} = i\Delta^{ij}.
\label{cnum}
\end{equation}
The higher-rank correlation tensors $W_{(n)}^{i_1\dots i_n}$ are then fixed by $W_{(2)}^{ij}$ through Wick's theorem. However for more general interacting theories the higher-rank correlation tensors will be necessary to specify the theory.

It is now clear that the entanglement entropy of $\rho$ relative to a region $\mathcal R$, $S(\mathcal R)$, can be expressed in terms of $W_{(2)}^{ij}$ for a free theory. But for an interacting theory, we would expect the higher correlators $W_{(n)}^{i_1\dots i_n}$ to contribute as well
\begin{equation}
    S(\mathcal R) = f(W_{(2)}, W_{(3)}, \dots).
\end{equation}

As a final remark, we note that not all theories are completely specified from the set of $n$-point functions of local operators. For example, in gauge theories non-local operators can exist such as line and surface operators \cite{Gukov:2014gja}.

\section{Entropy for the Gaussian Theory}\label{gaussianSee}
\subsection{Computation using Spacetime Correlators \texorpdfstring{$\Delta$ and $W$}{}}\label{freecorrelator}

In this subsection we review the main results in \cite{rssee}. We consider a free scalar field theory in a Gaussian state. That the theory is free means that the equations of motion are linear. As we have mentioned already,  the spacetime commutator in this case is a c-number.

The two field theory correlators that we need to define the entropy are
\begin{equation}
i\Delta(x,x')=[\phi(x),\phi(x')],
\label{delta}
\end{equation}
and\footnote{Since we are considering a free theory, the only $W_{(n)}$ we will need is $W_{(2)}$. In this subsection we will drop the  subscript and refer to $W_{(2)}$ as simply $W$.}
\begin{equation}
    W(x,x')=\langle \phi(x)\phi(x')\rangle
    \label{wightman}.
\end{equation}
As we have seen in \eqref{cnum}, $\Delta$ can be obtained from the anti-symmetric part of $W$. Other useful relations between these two are
\begin{equation}
    \Delta=2\,\text{Im}(W),\indent W=R+i\Delta/2,    
\end{equation}
where $R$ is real and symmetric. 
The field theory problem can be divided into a series of calculations each involving a single degree of freedom $q$ and $p$ (see \cite{rssee}). It suffices to derive the entropy for one degree of freedom and the full entropy becomes a sum of this quantity for all degrees of freedom. The most generic Gaussian density-matrix in a $q$-basis is
\begin{equation}
\rho(q,q')\equiv \langle q|\rho|q'\rangle=N ~e^{-A/2(q^2+q'^2)+iB/2(q^2-q'^2)-C/2(q-q')^2},\label{rhog}
\end{equation}
where $A,\,B,\,C$ are constant parameters and $N$ is a normalization constant that is fixed by the condition $\text{tr}\rho=1$. For \eqref{rhog} we get $N=\sqrt{A/\pi}$. We will work with \eqref{rhog} in this and the next subsection. 

Since the entropy $S$ is dimensionless and invariant under unitary transformations ($S(\rho)=S(U\rho U^\dagger)$), we expect it to depend on the symplectic invariant $\langle qq\rangle \langle pp\rangle-(\text{Re}\langle qp\rangle)^2=\det R/\det\Delta$. 
 The strategy will be to obtain $\Delta$ and $R$ in terms of $A,\, B,\, C$ and then relate $ \text{Spec}\{\Delta^{-1} R\}=\{i\sigma,-i\sigma\}$ to these parameters, thereby getting $S(\sigma)$. Also notice that $\sigma^2=\det R/\det \Delta$. In terms of $q$ and $p$, $\Delta$ and $R$ are
\begin{equation}
\Delta= 2 \text{Im}
  \begin{pmatrix} \langle qq\rangle & \langle qp\rangle \\\ \langle pq \rangle & \langle pp \rangle 
 \end{pmatrix}=  \begin{pmatrix} 0&1 \\ -1&0 \end{pmatrix},
\end{equation}
\begin{equation}
R= \text{Re}
  \begin{pmatrix} \langle qq\rangle & \langle qp\rangle \\\ \langle pq \rangle & \langle pp \rangle 
 \end{pmatrix}=  \begin{pmatrix} \langle qq\rangle & \text{Re}\langle qp\rangle \\\ \text{Re}\langle pq \rangle & \langle pp \rangle  \end{pmatrix}.
\end{equation} 
We need the correlators $ \langle qq\rangle,\, \langle qp\rangle,\,  \langle pq\rangle,$ and $\langle pp\rangle$ to obtain $\det R /\det\Delta$ in terms of $A$ and $C$. $B$ drops out of this expression \cite{rssee}, so the entropy will not depend on it.

We carry out the expectation value computations with $B$ set to zero in Appendix~\ref{correlationsg}. The results are
\begin{equation}
    \langle \hat{q}\hat{q}\rangle=1/(2A),\indent \langle \hat{q}\hat{p}\rangle=i/2, \indent    \langle \hat{p}\hat{q}\rangle= -i/2,\indent \langle \hat{p}\hat{p}\rangle=A/2+C.
    \label{corrsg}
\end{equation}
Putting the expressions above together we have 
\begin{equation}
\sigma^2=\langle qq\rangle \langle pp\rangle-(\text{Re}\langle qp\rangle)^2=1/(2A)(A/2+C)-0=1/4+C/(2A)
\label{sigmaca}
\end{equation}
which gives us the relation we need to switch between $\sigma$ and the parameters $A$ and $C$. At this stage, a result from \cite{bombelli} can be used where the entropy is obtained in terms of $\mu=\frac{\sqrt{1+2C/A}-1}{\sqrt{1+2C/A}+1}$ as
\begin{equation}
    S=-\frac{\mu\log\mu+(1-\mu)\log(1-\mu)}{1-\mu} .
    \label{smu}
\end{equation}
Using \eqref{sigmaca} we can rewrite \eqref{smu} in terms of $\sigma$ as
\begin{equation}
    S=(\sigma+1/2)\log(\sigma+1/2)-(\sigma-1/2)\log(\sigma-1/2).
    \label{ssigma}
\end{equation}
From here, further simplifying algebra is used to express \eqref{ssigma} in terms of the eigenvalues of $\Delta^{-1}W$ which are $\text{Spec}(\Delta^{-1}W)=\{i\omega_+,i\omega_-\}=\text{Spec}(\Delta^{-1}R+i/2)=\{i(1/2+\sigma),i(1/2-\sigma)\}$ as
\begin{equation}
    S=\omega_+\log\omega_+-\omega_-\log\omega_- .
\end{equation}

So far we have been considering the contribution of one degree of freedom to the entropy. The final step in our review is to include the contribution of all the degrees of freedom of the scalar field. As mentioned earlier, this can be done by summing over the contributions of each individual degree of freedom. Labelling now the eigenvalues as $\lambda$, where
\begin{equation}
    W v=i\lambda\Delta v,\indent \Delta v\neq 0,
\label{ss}
\end{equation}
our final expression for the entropy of a Gaussian scalar field is
\begin{equation}
S=\sum_{\lambda} \lambda\ln |\lambda|
\label{s4}.
\end{equation}
For some applications of~\eqref{ss}--\eqref{s4}, %\crefrange{ss}{s4},
where the conventional spatial area law scaling with the UV cutoff is obtained, see \cite{yas1,yas2}.
\subsection{Computation using Replica Trick}
\label{freereplica}
 In this subsection, we derive the result of the previous subsection using the replica trick with the density matrix \eqref{rhog}. See also \cite{Koksma}. The replica trick \cite{replicadef} is
 \begin{equation}
     S = -\tr\left(\rho\log\rho\right) = -\lim_{n\rightarrow 1}\frac{\partial}{\partial n}\tr\left(\rho^n\right),
\label{replica}
 \end{equation}
 where $n$ is analytically continued to $1$. We must therefore compute $\tr\left(\rho^n\right)$. The trace is by definition given by (with the notation $q_{n+1} = q_1$)
 \begin{eqnarray}
  \tr\left(\rho^n\right)&=&N^n\int dq_1...dq_n\,\rho(q_1,q_2)\rho(q_2,q_3)...\rho(q_n,q_1)\nonumber\\
  &=& N^n \int d^nq \exp{\left(\sum^{n}_{i=1}\left[-\frac{A}{2}(q_i^2+q_{i+1}^2)-\frac{C}{2}(q_i-q_{i+1})^2\right]\right)}\nonumber\\
  &=& N^n \int d^nq \exp{\left(-(A+C)\sum^{n}_{i=1}q_i^2+C\sum^{n}_{i=1}q_iq_{i+1}\right)}.
  \label{rhogn}
 \end{eqnarray}
 We can rewrite \eqref{rhogn} more conveniently as 
 
 \begin{equation}
    \tr\left(\rho^n\right)= N^n\int d^nq\, \exp{\left(-\beta (q_1-\mu q_2)^2-...-\beta(q_n-\mu q_1)^2\right)},
    \label{rhognb}
 \end{equation}
 where $\mu=\frac{\sqrt{1+2C/A}-1}{\sqrt{1+2C/A}+1}$ as before, and $\beta=\frac{1}{2} \left(A\sqrt{1+2 C/A}+A+C\right)$. In terms of these new parameters the $n^{th}$ power of the normalization constant is $N^n=\left(\frac{A}{\pi}\right)^{n/2}=|1-\mu|^n\left(\frac{\beta}{\pi}\right)^{n/2}$. 
 
 We next make the change of variables $y_i\equiv q_i-\mu q_{i+1}$. In terms of the Jacobian matrix $J$, $dq_i=\frac{\partial q_i}{\partial y_j}dy_j=(J^{-1})_{ij}dy_j$, where $J_{ij}=\frac{\partial y_i}{\partial q_j}=\delta_{ij}-\mu\delta_{i+1,j}$. For the full measure we have $d^nq=|J|^{-1}d^ny$, where $|J|=|1-\mu^n|$ is the determinant.
With the new variables, \eqref{rhognb} becomes
 \begin{eqnarray}
    \tr\left(\rho^n\right)&=& N^n |1-\mu^n|^{-1}\int dy_1 e^{-\beta y_1^2}\int dy_2 e^{-\beta y_2^2}\int...\int dy_n e^{-\beta y_n^2}\nonumber\\
   &=& N^n |1-\mu^n|^{-1}\left(\frac{\pi}{\beta}\right)^{n/2}=\frac{|1-\mu|^n}{|1-\mu^n|},
   \label{rhogny}
 \end{eqnarray}
 where we have substituted in the value of $N^n$ in the last step.
 
 Finally, we insert \eqref{rhogny} into \eqref{replica} and take the limit to obtain
 \begin{eqnarray}
  S&=&   -\lim_{n\rightarrow 1}\frac{\partial}{\partial n}\tr\left(\rho^n\right)\nonumber\\
  &=&    -\lim_{n\rightarrow 1}\frac{\partial}{\partial n}\left(\frac{|1-\mu|^n}{|1-\mu^n|}\right)\nonumber\\
  &=&-\lim_{n\rightarrow 1} \frac{\log (1-\mu ) (1-\mu )^n}{1-\mu ^n}+\frac{\log (\mu ) \mu ^n (1-\mu )^n}{\left(1-\mu ^n\right)^2}\nonumber\\
  &=& -\frac{\mu\log\mu+(1-\mu)\log(1-\mu)}{1-\mu}
  \label{replicasmu}.
 \end{eqnarray}
 The entropy expression \eqref{replicasmu} is exactly that of \eqref{smu}, and from \eqref{smu} we can follow the same procedure as in Section \ref{freecorrelator} that leads to \eqref{s4}.
 
\section{Entropy for a Perturbed Theory}
\label{quartic}
As a first extension of previous work in the Gaussian theory, we consider a  non-Gaussian theory. As in the previous section, we will compute the entropy independently via both the correlators and the replica trick.
  \subsection{Computation using Spacetime Correlators \texorpdfstring{$\Delta$ and $W$}{}}
  \label{intcorrelator}
 In this subsection we conjecture that to first order in perturbation theory, the entropy formula in the Gaussian case stays the same, but with $W$ and $\Delta$ replaced by those of the non-Gaussian theory. We will later return to the justification of this conjecture.\footnote{In Section \ref{intreplica} we will carry out an independent calculation of the entropy using the replica trick. In this replica trick calculation no assumptions are made regarding the form of the entropy and the result therein will be used to compare to the result in the present subsection.} With this assumption, we carry out the correlator calculations of Section \ref{freecorrelator} to first order in perturbation theory. 
 
We consider the density matrix (in a block)

\begin{equation}
\label{eq:rho}
 \rho_{qq'}=\langle q|\rho|q'\rangle=N e^{-A/2(q^2+q'^2)-C/2 (q-q')^2-\left(\lambda_1 \frac{q^4+q'^4}{2}+\lambda_2(q^3q'+qq'^3)+\lambda_3 q^2q'^2\right)} ,
\end{equation}
which is the most generic (symmetric in $q$ and $q'$) quartic perturbation to a Gaussian density matrix in a block.
By imposing that this density matrix is normalized (to first order in $\lambda_i$) we get
\begin{equation}
\label{eq:normalization}
    N \simeq \sqrt{\frac{A}{\pi}}  \left( 1 + \frac{3}{4A^2}(\lambda_1 + 2\lambda_2 +\lambda_3) \right).
\end{equation}

With this matrix we obtain the following correlators (always to first order in $\lambda_i$)
\begin{equation}
    \label{eq:corrqn}
    \langle \hat{q}^n\rangle \simeq \langle \hat{q}^n\rangle_g + (\lambda_1+2\lambda_2+\lambda_3) \left(
    \frac{3}{4A^2} \langle \hat{q}^n\rangle_g - \langle \hat{q}^{n+4}\rangle_g
    \right) ,
\end{equation}
where the subscript $g$ refers to the correlators in the Gaussian case. We can express the remaining correlators in terms of $\langle \hat{q}^n\rangle$ as
\begin{equation}
 \langle \hat{p}^2\rangle \simeq (A+C)+(6\lambda_1+6\lambda_2+2\lambda_3-A^2) \langle \hat{q}^2\rangle
 - 4A(\lambda_1+2\lambda_2+\lambda_3) \langle \hat{q}^4  \rangle
\end{equation}
and
\begin{equation}
     \langle \hat{q}\hat{p}\rangle
    = \overline{\langle \hat{p}\hat{q}\rangle} \simeq i\left(A\langle \hat{q}^2\rangle 
     + 2(\lambda_1+2\lambda_2+\lambda_3) \langle\hat{q}^4\rangle \right) = i/2 \,.
\end{equation}
The details of the computations of the above correlators can be found in Appendix~\ref{corrsng}.

The computation of $\sigma^2$ (recall from equation~\eqref{sigmaca}) reduces to
\begin{eqnarray}
\sigma^2&=&\langle \hat{q}^2\rangle \langle \hat{p}^2\rangle-0 \nonumber
\simeq(A+C) \langle \hat{q}^2\rangle-A^2\langle \hat{q}^2\rangle\langle \hat{q}^2\rangle \nonumber \\
 && + 2\langle \hat{q}^2\rangle\langle \hat{q}^2\rangle(3\lambda_1 + 3\lambda_2 + \lambda_3)
 - 4 A \langle \hat{q}^4 \rangle \langle \hat{q}^2\rangle (\lambda_1 + 2\lambda_2 + \lambda_3),
\end{eqnarray}
and using \eqref{eq:corrqn} we get
\begin{eqnarray}
\sigma^2&\simeq&(A+C) \left( \langle \hat{q}^2\rangle_g + (\lambda_1+2\lambda_2+\lambda_3) \left(
    \frac{3}{4A^2} \langle \hat{q}^2\rangle_g - \langle \hat{q}^{6}\rangle_g
    \right) \right) \nonumber \\
 && -A^2 \left( \langle \hat{q}^2\rangle_g \langle \hat{q}^2\rangle_g 
 + 2 \langle \hat{q}^2\rangle_g (\lambda_1+2\lambda_2+\lambda_3) \left(
    \frac{3}{4A^2} \langle \hat{q}^2\rangle_g - \langle \hat{q}^{6}\rangle_g
    \right) \right) \nonumber \\
 && + 2\langle \hat{q}^2\rangle_g \langle \hat{q}^2\rangle_g (3\lambda_1 + 3\lambda_2 + \lambda_3)
 - 4 A \langle \hat{q}^4 \rangle_g \langle \hat{q}^2\rangle_g (\lambda_1 + 2\lambda_2 + \lambda_3) \nonumber \\
 &=& \frac{1}{4} + \frac{C}{2A} - \frac{3(A+C)}{2A^3}(\lambda_1 + 2\lambda_2 + \lambda_3)
 +\frac{1}{2A^2}(3\lambda_1 + 3\lambda_2 + \lambda_3) .
\end{eqnarray}

For later comparison we will compute the following quantity
\begin{gather}\label{eq:mucorrelation}
    \begin{aligned}
        \mu_{\mathrm{correlation}} &= \frac{2\sigma - 1}{2\sigma + 1}\\
        &\simeq \mu + \frac{3\mu}{(\mu+1)(\mu-1)^3}\lambda_1 + \frac{3(\mu+1)}{2(\mu-1)^3\beta^2}\lambda_2 + \frac{1+\mu+\mu^2}{(\mu+1)(\mu-1)^3\beta^2}\lambda_3
    \end{aligned}
\end{gather}
to first order in perturbation theory. This is nothing but the quantity appearing in the Gaussian spacetime entanglement entropy formula, computed perturbatively for a non-Gaussian theory. We will investigate whether the entanglement entropy for the non-Gaussian theory can be obtained from the formula \eqref{smu}, by replacing $\mu$ with $\mu_{\mathrm{correlation}}$.

 \subsection{Computation using Replica Trick}
 \label{intreplica}
 
 We will now compute the entanglement of the perturbed state
 \begin{equation}
 \rho_{qq'}=\langle q|\rho|q'\rangle=N e^{-A/2(q^2+q'^2)-C/2 (q-q')^2-\left(\lambda_1 \frac{q^4+q'^4}{2}+\lambda_2(q^3q'+qq'^3)+\lambda_3 q^2q'^2\right)}
 \end{equation}
 using the replica trick and perturbation theory. The trace of $\rho^n$ can be parameterized in the following way
 \begin{gather} \label{eq:rhoNperturbed}
     \begin{aligned}
        \tr\left(\rho^n\right) &= N^n\int d^nq\exp\left(-\frac 12q^TMq + q^\alpha q^\beta q^\gamma q^\delta T_{\alpha\beta\gamma\delta}\right), \\
        &= N^n\int d^nq\exp\left(-\frac 12q^TMq\right)\sum_{i=0}^\infty\frac 1{i!} (q^\alpha q^\beta q^\gamma q^\delta T_{\alpha\beta\gamma\delta})^i,
     \end{aligned}
 \end{gather}
where the quartic perturbation tensor is given by
\begin{equation}\label{eq:T_tensor}
    T_{ijk\ell}=-\lambda_1\delta_{ij}\delta_{jk}\delta_{k\ell}-\lambda_2\delta_{ij}\delta_{jk}(\delta_{k+1,\ell}+\delta_{k-1,\ell})-\lambda_3\delta_{ij}\delta_{j+1,k}\delta_{k\ell},
\end{equation}
and the quadratic coefficient matrix is given by
\begin{gather}\label{eq:M_matrix_components}
    \begin{aligned}
        M_{ij} &=    2(A+C)\delta_{ij} - C\left(\delta_{i,j+1} + \delta_{i+1, j}\right),\\
               &=    2\beta(1+\mu^2)\delta_{ij} - 2\beta\mu\left(\delta_{i,j+1} + \delta_{i+1, j}\right),
    \end{aligned}
\end{gather}
with periodic convention of the indices $i = n+1 = 1$ (or in other words, the Kronecker delta is over $\mathbb Z_n$ and indices defined modulo $n$). In the second line we have used $A=\beta(1-\mu)^2$ and $C = 2\beta\mu$.

It is useful to think of the above as the partition function of an interacting $n$-particle system and define the following non-interacting correlation functions
 \begin{equation}
     \llangle \mathcal O \rrangle \equiv \frac 1{Z_0}\int d^nq\exp\left(-\frac 12q^TMq\right)\mathcal O(q).
 \end{equation}
The Gaussian partition function is given by
 \begin{equation}
     Z_0 = \int d^nq\exp\left(-\frac 12q^TMq\right) = \sqrt{\frac{(2\pi)^n}{\det M}}=\frac{\left(\sqrt{\frac{\pi}{\beta}}\right)^n}{|1-\mu^n|},
 \end{equation}
 where in the last expression above we used \eqref{detM}.
The equation \eqref{eq:rhoNperturbed} can now be expressed in terms of familiar perturbation theory
 \begin{gather}\label{eq:RhoNPerturbation}
     \begin{aligned}
        \tr\left(\rho^n\right) &= N^nZ_0\sum_{i=0}^\infty\frac 1{i!} 
        \llangle q^{\alpha_1} q^{\beta_1} q^{\gamma_1}q^{\delta_1}\cdots q^{\alpha_i} q^{\beta_i} q^{\gamma_i} q^{\delta_i}\rrangle T_{\alpha_1\beta_1\gamma_1\delta_1}\cdots T_{\alpha_i\beta_i\gamma_i\delta_i}.
     \end{aligned}
 \end{gather}
 The $\mathcal O(T^m)$ correction requires a $4m$-point $\llangle\cdot\rrangle$ correlation function. In order to compute the $4m$-point correlator integrals, it is easier to first compute the integral
 \begin{equation}
     Z(J) = \frac 1{Z_0}\int d^nq\exp\left(-\frac 12q^TMq + J^Tq\right) = \exp\left(\frac 12J^TM^{-1}J\right),
 \end{equation}
 and then compute the correlators by differentiation
 \begin{gather}\label{eq:replicaWicks}
    \begin{aligned}
        \llangle q_{\alpha_1}\cdots q_{\alpha_{2m}}\rrangle &= \frac{\partial^{2m}}{\partial J_{\alpha_1}\cdots\partial J_{\alpha_{2m}}}Z(J)\Big|_{J=0}\\
        & = \frac 1{2^m m!}\sum_{\sigma\in S_{2m}}\left(M^{-1}\right)_{\alpha_{\sigma(1)\sigma(2)}}\cdots \left(M^{-1}\right)_{\alpha_{\sigma(2n-1)\sigma(2n)}}.
     \end{aligned}
 \end{gather}
 Note that this is nothing but Wick's theorem with $\llangle q_\alpha q_\beta\rrangle = \left(M^{-1}\right)_{\alpha\beta}$. We can however reduce this sum significantly as the terms have a lot of internal symmetry.
 Since $M$ is a symmetric matrix, any permutation of indices within a $M^{-1}$ factor will give an identical term. This leads to a multiplicity of 2 and thus in total $2^m$, if we use a convention that removes this redundancy. Furthermore there is an $m!$-fold multiplicity since each term is invariant under all permutations of the $M^{-1}$ factors (or more precisely, their indices). Therefore, we only need to consider a subset of permutations $G\subset S_{2m}$
 \begin{gather}\label{eq:CorrelatorsReduced}
    \begin{aligned}
        \llangle q_{\alpha_1}\cdots q_{\alpha_{2m}}\rrangle
        &=\sum_{\sigma\in G}\left(M^{-1}\right)_{\alpha_{\sigma(1)\sigma(2)}}\cdots \left(M^{-1}\right)_{\alpha_{\sigma(2n-1)\sigma(2n)}}.
     \end{aligned}
 \end{gather}
 Here we have defined the quotient of groups
  \begin{equation}
     G = \frac{S_{2m}}{\left[\prod_{i=0}^{m-1} S^{(2i+1,2i+2)}_2\right]\times \left[\prod_{(i\neq j)=0}^{m-1} S^{(2i+1, 2j+1)}_{p,2}\right]},
 \end{equation}
 where $S_2^{(i,j)}\subset S_{2m}$ is the subgroup of swaps between the $i$'th and $j$'th elements only and $S_{p,2}^{(i,j)}\subset S_{2m}$ (for odd $i\neq j$) are pairwise swaps of the $(i, i+1)$'th elements with the $(j, j+1)$'th elements. All these subgroups are isomorphic to $S_2$. The number of inequivalent permutations is thus $|G| = \frac{(2m)!}{2^m m!}$. For example for a 4-point function ($2m=4$), there are $3$ different permutations.

As computed in Appendix~\ref{appendix:M_inverse}, the inverse of $M$ can be expressed in different ways
\begin{equation}
    M^{-1}_{ij}=\sum_{k=1}^n\frac{\mu^{(k-i)\Mod{n} + (k-j)\Mod{n}}}{2(1-\mu^n)^2\beta} = \frac 1n\sum_{x=0}^{n-1}\frac {e^{i\frac {2\pi}n x(j-j')}}{2\beta(1+\mu^2) - 4\beta\mu\cos\left[\frac {2\pi}n x\right]}.
\end{equation}
This matrix is dense, but for our purposes we only need the diagonal
\begin{equation}
    M^{-1}_{ii}=\sum_{k=0}^{n-1}\frac{\mu^{2k}}{2(1-\mu^n)^2\beta}=\frac{(\mu^{2n}-1)}{2(1-\mu^n)^2\beta(\mu^2-1)},
\end{equation}
and the next-to-diagonal elements
\begin{equation}
    M^{-1}_{ii+1}=M^{-1}_{ii-1}=\frac{\sum_{k=0}^{n-2}\left(\mu^{2k+1}+\mu^{n-1}\right)}{2(1-\mu^n)^2\beta}=\frac{(\mu^n+\mu^2)}{2\mu\beta(\mu^2-1)(\mu^n-1)}.
\end{equation}
Here we have used that $M_{ii}=M_{11}$ and $M_{ii+1} = M_{12}$ for any $i$.
The determinant is given by
\begin{equation}
    \det M=2^n(1-(-\mu)^n)^2\beta^n.
\label{detM}
\end{equation}

Substituting \eqref{eq:CorrelatorsReduced} and \eqref{eq:T_tensor} into \eqref{eq:RhoNPerturbation}, to first order in $T$ we get
\begin{gather}
    \begin{aligned}
         \tr\left(\rho^n\right) &= N^nZ_0\left(1 + T_{ijk\ell}\left[M^{-1}_{ij}M^{-1}_{k\ell}+M^{-1}_{ik}M^{-1}_{j\ell}+M^{-1}_{i\ell}M^{-1}_{jk}\right]\right) + \mathcal O(T^2)\\
        &= N^nZ_0\left(1-n\left[(3\lambda_1+\lambda_3)(M^{-1}_{11})^2+6\lambda_2 M^{-1}_{12}M^{-1}_{12}+2\lambda_3 (M^{-1}_{12})^2\right]\right) + \mathcal O(\lambda^2)
    \end{aligned}
\end{gather}
Since we know the explicit $n$-dependence of $\tr\left(\rho^n\right)$, we can find the entropy using the replica trick from equation \eqref{replica}
\begin{gather}\label{eq:EEnongaussian}
    \begin{aligned}
        S=&-\frac{\mu\log\mu+(1-\mu)\log(1-\mu)}{1-\mu} - \frac{3\mu\log\mu}{(\mu+1)(\mu-1)^5\beta^2}\lambda_1\\&\qquad - \frac{3(\mu+1)\log\mu}{2(\mu-1)^5\beta^2}\lambda_2 - \frac{3(1+\mu+\mu^2)\log\mu}{(\mu+1)(\mu-1)^5\beta^2}\lambda_3 + \mathcal O(\lambda^2).
    \end{aligned}
\end{gather}
This is the entropy of the non-Gaussian theory up to first order in perturbation theory. We would like to investigate whether there exists a $\mu_{\mathrm{replica}}$ such that the entanglement entropy can be expressed in the following way
\begin{equation}\label{eq:EEnongaussianSpacetimeForm}
    S=-\frac{\mu_{\mathrm{replica}}\log\mu_{\mathrm{replica}}+(1-\mu_{\mathrm{replica}})\log(1-\mu_{\mathrm{replica}})}{1-\mu_{\mathrm{replica}}}.
\end{equation}
We can express this quantity perturbatively as 
\begin{equation}
    \mu_{\mathrm{replica}} = \mu + \sum_{i=1}^3\delta\mu_i\lambda_i + \mathcal O(\lambda^2).
\end{equation}
Inserting this into \eqref{eq:EEnongaussianSpacetimeForm} and expanding to first order we get
\begin{equation}
    S = -\frac{\mu\log\mu+(1-\mu)\log(1-\mu)}{1-\mu}  - \frac{\log\mu}{(\mu-1)^2}\sum_{i=1}^3\delta\mu_i\lambda_i + \mathcal O(\lambda^2).
\end{equation}
By setting this equal to equation \eqref{eq:EEnongaussian} and solving for $\delta\mu_i$ we find the following solutions
\begin{equation}
    \delta\mu_1 = \frac{3\mu}{\beta^2(\mu+1)(\mu-1)^3}, \quad \delta\mu_2 = \frac{3(\mu+1)}{2\beta^2(\mu-1)^3}, \quad \delta\mu_3 = \frac{1+\mu+\mu^2}{\beta^2(\mu+1)(\mu-1)^3}.
\end{equation}
We can thus parametrize the entanglement entropy in terms of the parameter
\begin{equation}
    \mu_{\mathrm{replica}} \simeq \mu + \frac{3\mu}{\beta^2(\mu+1)(\mu-1)^3}\lambda_1 + \frac{3(\mu+1)}{2\beta^2(\mu-1)^3}\lambda_2 + \frac{1+\mu+\mu^2}{\beta^2(\mu+1)(\mu-1)^3}\lambda_3.
\end{equation}
By comparing this to equation \eqref{eq:mucorrelation}, we see that
\begin{equation}
    \mu_{\mathrm{replica}} = \mu_{\mathrm{correlation}}.
\end{equation}
In other words $\mu_{\mathrm{replica}}$, and thus also the entanglement entropy, is actually a spacetime quantity and can be computed from spacetime correlators. The entanglement entropy formula \eqref{smu} appears to still hold beyond the Gaussian theory (where Wick's theorem holds). 

\section{Generalization to Arbitrary Perturbations and Higher Orders}
Our result above, for the first order contribution to the entanglement entropy, continues to hold for arbitrary perturbations. Namely, the first order (Gaussian or non-Gaussian) perturbation of the entanglement entropy around a Gaussian state (bosonic or fermionic, and involving an  arbitrary number of degrees of freedom) is completely captured by the two-point correlation function of the perturbed state. In this section we prove a general theorem stating this, and we discuss some of its implications.

\subsection{Proof of the General Case}
We consider a bosonic\footnote{While we focus on the bosonic case, our derivation applies analogously to fermions by treating them in a parallel way, as reviewed in \cite{hackl2020bosonic}.} system with $N$ degrees of freedom, classical phasespace $V$ and separable Hilbert space $\mathcal{H}_V$. We can always choose a basis of linear observables\footnote{They are called linear observables, because classical $q_i: V\to\mathbb{R}$ and $p_i: V\to\mathbb{R}$ are linear maps on phasespace.}
\begin{align}
    \hat{\xi}=(\hat \xi^1,\hat\xi^2,\dots, \hat\xi^{2N})\equiv(\hat{q}_1,\hat{p}_1,\dots,\hat{q}_N,\hat{p}_N),
\end{align}
which satisfy the canonical commutation relations\footnote{For fermions, $\hat{\xi}^a$ represents Majorana modes that satisfy canonical anticommutation relations, \ie $\{\hat{\xi}^a,\hat{\xi}^b\}=G^{ab}$ with positive definite bilinear form $G^{ab}$.}
\begin{align}
    [\hat{\xi}^a,\hat{\xi}^b]=\ii\Omega^{ab},
\end{align}
where $a,b=1,\dots,2N$ and
\begin{equation}
    \Omega = \bigoplus^N_{i=1}\begin{pmatrix}
    0 & 1\\
    -1 & 0
    \end{pmatrix}.
\end{equation}
In the context of field theory, we may refer to $\hat{q}_i$ and $\hat{p}_i$ rather as $\hat{\varphi}_i$ and $\hat{\pi}_i$, \ie as field operators and their conjugate momenta. When taking the continuum limit, these operators become operator-valued distributions, as commonly considered in algebraic quantum field theory.

A phase space decomposition $V=A\oplus B$ into subsystems $A$ and $B$ induces a Hilbert space decomposition $\mathcal{H}_V=\mathcal{H}_A\otimes\mathcal{H}_B$ with operators $\hat{\xi}^a_A$ of the form $\mathcal{O}_A\otimes \id_B$, \ie they only probe the state from the perspective of the subsystem $A$ which could represent a causal diamond in spacetime.

Let us consider a one-parameter family $\ket{\psi_\epsilon}$ of pure quantum states in $\mathcal{H}_V$, where we require $\ket{\psi_0}$ to be a Gaussian state. This induces a one-parameter family
\begin{align}
    \rho_\epsilon=\tr_{\mathcal{H}_B}\ket{\psi_\epsilon}\bra{\psi_\epsilon}   
\end{align}
of possibly mixed states by tracing out the degrees of freedom of $\mathcal{H}_B$. By construction, $\rho_0$ is Gaussian and can therefore be written as $\rho_0=e^{-\hat{H}_A}/Z$ with $Z=\tr e^{-\hat{H}_A}$, where $\hat{H}_A$ is known as the ``modular Hamiltonian'', which is quadratic for Gaussian states. This means there exists a symmetric, real bilinear form $h_{ab}$ with\footnote{For fermions, we would have $\hat{H}_A=\frac{i}{2}h_{ab}\hat{\xi}^a_A\hat{\xi}^b_A$, where $h_{ab}$ would be an antisymmetric and real bilinear form.}
	\begin{align}
		\hat{H}_A=\frac{1}{2}h_{ab}\,\hat{\xi}^a_A\hat{\xi}^b_A\,.
	\end{align}
The entanglement entropy is given as a function $S_\epsilon=-\tr_{\mathcal{H}_A}(\rho_\epsilon\log\rho_\epsilon)$, which can be computed at $\epsilon=0$ from the two-point correlation function via \eqref{ss}--\eqref{s4}.\footnote{Compared to our earlier perturbations in terms of  $\lambda$'s, one could think of $\epsilon$ as a perturbation along a particular direction $\hat r$ in parameter space: $(\lambda_1,\lambda_2, \lambda_3)=\epsilon\hat r$. This is essentially writing the parameters in spherical coordinates.} The first law of entanglement entropy \cite{bhattacharya2013thermodynamical,fl1,fl2} (see a brief derivation in the next subsection) states that we have
	\begin{align}
		S_\epsilon=S_0+\epsilon \frac{d}{d\epsilon}\bigg|_{\epsilon=0}\!\!\tr(\hat{H}_A\,\rho_\epsilon)+\mathcal{O}(\epsilon^2)\,.
	\end{align}
Since $\hat{H}_A$ is  quadratic  (for both bosons and fermions), it only probes the two-point correlation function of $\rho_\epsilon$. It therefore does not matter if $\ket{\psi_\epsilon}$ is perturbed in a Gaussian or non-Gaussian way, as the entanglement entropy at linear order will only be sensitive to the two-point correlation function 
\begin{align}
    C^{ab}_A(\epsilon)=\mathrm{tr}(\hat{\xi}^a_A\hat{\xi}^b_A\rho_\epsilon)
\end{align}
of the state $\rho_\epsilon$ which cannot be distinguished from a Gaussian state. We find\footnote{For fermions, we find almost the same expression, but the RHS of \eqref{eq:Sprime} will contain an additional $\ii$.}
\begin{align}
    S'_0=\frac{d}{d\epsilon}\braket{\hat{H}_A}\big|_{\epsilon=0}=\frac{1}{2}h_{ab}\frac{d}{d\epsilon}C^{ab}_A(\epsilon)\big|_{\epsilon=0}\,.\label{eq:Sprime}
\end{align}
Put differently, for any highly non-Gaussian family $\rho_\epsilon$, we could define a Gaussian one-parameter family $\tilde{\rho}_\epsilon$, such that $\mathrm{tr}(\hat{\xi}_A^a\hat{\xi}_A^b\rho_\epsilon)=\tr(\hat{\xi}_A^a\hat{\xi}_A^b\tilde{\rho}_\epsilon)$. The entanglement entropy would not be able to distinguish between the two at linear order around $\rho_0$. This last point ensures that the formula for the linear perturbation of the entanglement entropy will be the same as expanding formulas \eqref{ss}--\eqref{s4} for Gaussian states, which explains the finding of the previous sections.\footnote{For fermions, there is an analogous formula derived in\ \cite{peschel2003calculation}. There is also a unified framework of computing the entanglement entropy for both bosons and fermions in terms of the so-called linear complex structure\ \cite{bianchi2015entanglement,vidmar2017entanglement,hackl2018aspects}.}

\begin{figure}
    \centering
    \begin{tikzpicture}[scale=1.5]
    
    \begin{scope}[yshift=-4cm]
    \draw[every to/.style={out=-20,in=160,relative},red,very thick] ($.5*(5.5,5)-.1*(-4.9,1.8)+.5*(5.5,3.2)$) to ($.5*(10.4,3.2)+.1*(-4.9,1.8)+.5*(10.4,5)$) node[right]{mixed Gaussian states on $\mathcal{H}_A$};
    \end{scope}
    
    \draw[very thick,red] (5.99,-.08) -- (5.99,1.72);
    \draw[very thick,red] (9.91,.28) -- (9.91, 2.08);
    
    \draw[thick] (10.4,2.8) -- node[right]{mixed states on $\mathcal{H}_A$} (10.4,1) (5.5,1) -- (5.5,-.8) (6.5,2.43) -- (6.5,.63);
    \manifold[black,thick]{5.5,-.8}{10.4,1}
    \draw[thick]  (9.43,1.35) -- (9.43,-.45);
    \manifold[black,thick]{5.5,1}{10.4,2.8}

    \begin{scope}[yshift=-2.2cm]
    \draw[every to/.style={out=-20,in=160,relative},red,very thick] ($.5*(5.5,5)-.1*(-4.9,1.8)+.5*(5.5,3.2)$) to ($.5*(10.4,3.2)+.1*(-4.9,1.8)+.5*(10.4,5)$);
    \end{scope}

    \begin{scope}[yshift=-1.1cm]
    \draw (10.4,5) node[right]{projective Hilbert space $\mathcal{P}(\mathcal{H}_V)$};
    \manifold[black,thick,fill=white,fill opacity=0.95]{5.5,3.2}{10.4,5}
    \draw[every to/.style={out=-20,in=160,relative},red,very thick] ($.5*(5.5,5)-.1*(-4.9,1.8)+.5*(5.5,3.2)$) to ($.5*(10.4,3.2)+.1*(-4.9,1.8)+.5*(10.4,5)$) node[right]{pure Gaussian states $\mathcal{M}$};
    
    \draw[very thick,black!40!green,->] plot [smooth] coordinates {(6.7,3.4) (7.2,3.6) (7.95, 4.1) (8.25,4.4) (9,4.7)};
    \draw[black!40!green] (9,4.7) node[right]{$\psi_{\epsilon}$};
    \draw[dotted] (8.2,4.5) --(8.27,4.19);
    \draw[thick,->] (7.95, 4.1) -- (8.2,4.5) node[left,scale=.8]{$\delta\psi_0$};
    
    \begin{scope}[yshift=-2.4cm]
    \draw[very thick,black!40!green,->] plot [smooth] coordinates {(6.7,3.1) (7.2,3.4) (7.95, 4.1) (8.25,4.4) (9,4.7)};
    \draw[black!40!green] (9,4.7) node[right]{$\rho_{\epsilon}$};
    \draw[dotted] (8.2,4.5) --(8.27,4.19);
    \draw[thick,->] (7.95, 4.1) -- (8.2,4.5) node[left,scale=.8]{$\delta\rho_0$};
    \draw[thick,->,purple] (7.95, 4.1) -- (8.27,4.19) node[below,xshift=8pt,scale=.8,yshift=-3pt]{$(\delta\rho_0)_{\parallel}$};
    \fill (7.95, 4.1) node[below]{$\rho_{0}$} circle (1.5pt);
    \end{scope}
    
    \draw[thick,->,purple] (7.95, 4.1) -- (8.27,4.19) node[below,xshift=8pt,scale=.8,yshift=-3pt]{$(\delta\psi_0)_{\parallel}$};
    \fill (7.95, 4.1) node[below]{$\psi_{0}$} circle (1.5pt);
    
    \end{scope}
    \end{tikzpicture}
    \caption{We sketch how the manifold of pure Gaussian states on $\mathcal{H}_V$ reduce to mixed states on $\mathcal{H}_A$. Both manifolds contain the submanifolds of Gaussian states (indicated in red). A trajectory $\psi_{\epsilon}$ of pure states that intersect the manifold of pure Gaussian states reduces to a trajectory of mixed states $\rho_{\epsilon}=\tr_B\ket{\psi_{\epsilon}}\bra{\psi_{\epsilon}}$. At the Gaussian state $\psi_0$ and its reduction to $\rho_0$ on $\mathcal{H}_A$, we can project its derivative $\delta\psi_0=\psi_0'$ and $\delta\rho_0=\rho_0'$ onto the Gaussian tangent space, which is fully characterized by the change of the two-point correlation function (see~\cite{hackl2020geometry}). The linear change of the entanglement entropy $\delta S_0=S'_0$ around $\psi_0$ only depends on this projected component $(\delta\psi_0)_{\parallel}$.}
    \label{fig:geometry-sketch}
\end{figure}
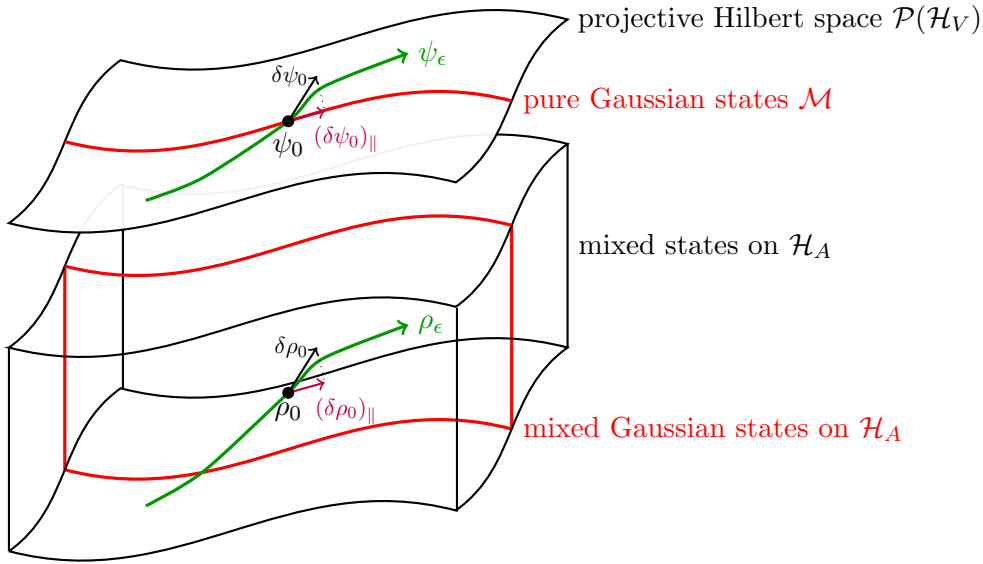

This result can also be interpreted geometrically as indicated in Figure~\ref{fig:geometry-sketch}. The manifold of pure Gaussian states $\mathcal{M}$ is a submanifold of the projective Hilbert space $\mathcal{P}(\mathcal{H}_V)$ consisting of all pure states. At a Gaussian state $\psi_0$, we have
\begin{align}
    \mathcal{T}_{\psi_0}\mathcal{M}\subset\mathcal{T}_{\psi_0}\mathcal{P}(\mathcal{H}_V)\,.
\end{align}
Due to the fact that these tangent spaces are equipped with a natural Hilbert space inner product enables us to decompose any linear perturbation $\delta\psi_0$ of a state $\psi_0$ into two pieces, namely
\begin{align}
    (\delta\psi_0)_{\parallel}\in\mathcal{T}_{\psi_0}\mathcal{M}\quad\text{and}\quad(\delta\psi_0)_{\perp}\in\mathcal{P}(\mathcal{H}_V)\,,
\end{align}
which are orthogonal to each other. One can show that $(\delta\psi_0)_{\parallel}$ captures the full change of the covariance matrix~\cite{hackl2020geometry}, while $(\delta\psi_0)_{\perp}$ only feels the change of higher order correlation functions.\footnote{If we allow for the change of the expectation values $z^a=\braket{\hat{\xi}^a}$, it will be part of our manifold $\mathcal{M}$ of squeezed coherent Gaussian states. In this case, $(\delta\psi_0)_{\parallel}$ will be fully determined by both the one- and two-point correlation functions.} In this picture, the entanglement entropy as a function of the pure state $\psi_{\epsilon}$ at $\psi_0$ only depends on the Gaussian perturbation $(\delta\psi_0)_{\parallel}=\frac{d}{d\epsilon}\psi_{\epsilon}|_{\epsilon=0}$.

In summary, we proved the observation of the previous section in full generality (for any non-Gaussian perturbation of a Gaussian state) using the properties of pure and mixed Gaussian states. While we focused on the bosonic case, we commented at the relevant places how we would arrive at the same conclusion for the linear perturbation of the entanglement entropy around fermionic Gaussian states.
	
\subsection{Analysis of Second Order}
We would like to compute the higher order perturbations of the entanglement entropy. We define $\rho_\epsilon=e^{-\hat{H}_\epsilon}$, where $\hat{H}_\epsilon$ is the modular Hamiltonian. We have
	\begin{align}
		S_\epsilon=-\tr(\rho_\epsilon\log{\rho_\epsilon})=S_0+\epsilon\, S'_0+\epsilon^2\,S''_0+\mathcal{O}(\epsilon^3)\,.
	\end{align}
As discussed in the previous subsection, the first order perturbation is well-known from the first law of entanglement entropy and given by
	\begin{align}
		S_\epsilon'=-\tr\left(\rho_\epsilon'\log{\rho_\epsilon}+\rho_\epsilon\rho_\epsilon^{-1}\rho'_\epsilon\right)=-\mathrm{tr}(\rho_\epsilon'\log{\rho_\epsilon})=\mathrm{tr}(\rho_\epsilon'\hat{H}_\epsilon)\,,
	\end{align}
where we used that $\tr(\rho_\epsilon')=\tr(\rho_\epsilon)'=0$ as the trace of a mixed state is constant and equal to $1$. Note that we did not need to worry about the ordering of the inverse of $\rho_\epsilon$, because the expression was inside a trace where we can use cyclicity. If $\rho_\epsilon$ were not invertible, we mean by $\rho_\epsilon^{-1}$ the Penrose-Moore pseudo-inverse, where we invert on the orthogonal complement of its kernel, without needing to change the resulting equations. As previously discussed, the first order perturbation $S'_0$ away from a Gaussian state (where $\hat{H}_0$ is quadratic) will only depend on the change of the two point function of the state $\rho_\epsilon$.
	
The second order perturbation can then be computed as
	\begin{align}
		S''_\epsilon=-\mathrm{tr}(\rho_\epsilon''\log\rho_\epsilon+\rho_\epsilon'\rho_\epsilon^{-1}\rho_\epsilon')\,.\label{Sepsilon2}
	\end{align}
We see that this perturbation consists of two pieces. The first one will be the second order change $\mathrm{tr}(\rho''\hat{H})$ of the expectation value of the modular Hamiltonian, which will again only depend on the change of the two-point function when perturbing a Gaussian state. The second piece is less straightforward and given by $-\mathrm{tr}(\rho_0^{'2}\rho_0^{-1})$. It is expected that this second piece will probe genuine non-Gaussian properties. Let us take a closer look at this term. Pulling out the derivative from $(\rho'_\epsilon)^2$ we get
	\begin{gather}
	    \begin{aligned}
	    (\rho'_\epsilon)^2 &= \frac{\partial}{\partial\epsilon}\left(\rho'_\epsilon\rho_\epsilon\right) - \rho''_\epsilon\rho_\epsilon\\
	    &= \frac{\partial}{\partial\epsilon}\frac{\partial}{\partial\sigma}\left(\rho_\sigma\rho_\epsilon\right)\bigg|_{\sigma=\epsilon} - \rho''_\epsilon\rho_\epsilon
	    \end{aligned}
	\end{gather}
The second term above has vanishing contribution to  $S''_0$:  
    \begin{equation}
        \frac{\partial^2}{\partial\epsilon^2}\tr\left(\rho_\epsilon\rho_0\rho_0^{-1}\right) = \frac{\partial^2}{\partial\epsilon^2}\tr\left(\rho_\epsilon\right) = 0.
    \end{equation}
The first term yields    
    \begin{equation}
        \tr\left((\rho_\epsilon')^2\rho_\epsilon^{-1}\right)\bigg|_{\epsilon=0} = \frac{\partial}{\partial\epsilon}\frac{\partial}{\partial\sigma}\left(\rho_\sigma\rho_\epsilon\rho_0^{-1}\right)\bigg|_{\sigma=\epsilon}\bigg|_{\epsilon=0}.
    \end{equation}
Since $\rho_\epsilon$ and $\rho_\sigma$ do not commute, unless $\epsilon=\sigma$, we have
    \begin{equation}
        A_{\sigma,\epsilon}\equiv \rho_\sigma\rho_\epsilon = e^{H_\epsilon + H_\sigma + \frac 12 \left[H_\epsilon, H_\sigma\right] + \dots}.
    \end{equation}
Note that $A_{\sigma, \epsilon}$ is not in general a density matrix, nor even Hermitian. We can however use a polar decomposition to decompose it into
    \begin{equation}
        A_{\sigma,\epsilon} = \rho_{\sigma,\epsilon}U_{\sigma,\epsilon},
    \end{equation}
a positive Hermitian operator $\rho_{\sigma,\epsilon}$ and a unitary operator $U_{\sigma,\epsilon}$. These operators are
    \begin{equation}
        \rho_{\sigma,\epsilon} = (A^\dagger A)^{\frac 12} = (\rho_{\epsilon}\rho_{\sigma}^2\rho_{\epsilon})^{\frac 12}, \qquad \text{and}\qquad U_{\sigma,\epsilon} =  A_{\sigma,\epsilon}\rho_{\sigma,\epsilon}^{-1}=\rho_{\sigma}\rho_{\epsilon}(\rho_{\epsilon}\rho_{\sigma}^2\rho_{\epsilon})^{-\frac 12}.
    \end{equation}
Here $\rho_{\sigma,\epsilon} = \exp(H_{\sigma,\epsilon})$ is a two-parameter family of states where $\rho_{0,0} = \rho_0^2$ is Gaussian, but $H_{\sigma,\epsilon}$ is in general some interacting Hamiltonian. Recall that\footnote{There is a very compact way to compute $\rho_0$ from the $2$-point correlation function derived in\ \cite{hackl2018aspects}. For this, it is best to express the two-point function in terms of the linear complex structure $J$ as in \cite{bianchi2015entanglement}. This gives $\rho_0=\exp{(-q_{ab}\hat{\xi}^a\hat{\xi}^b)}/Z$ where $q=-i\omega \,\mathrm{arccoth}(i J)$ for bosons and $q=i g\,\mathrm{arctanh}(i J)$ for fermions, where $\omega$ is the inverse symplectic form (commutation relations), $g$ the inverse metric (anti-commutation relations) and the equations should be understood as matrix equations. Note that this formula only applies to the Gaussian state $\rho_0=\rho_{0,0}$.} $\rho_0^{-1} = Z e^{-H_0}$, then
\begin{gather}
    \begin{aligned}
    \tr\left((\rho_\epsilon')^2\rho_\epsilon^{-1}\right)\bigg|_{\epsilon=0} &= Z\frac{\partial}{\partial\epsilon}\frac{\partial}{\partial\sigma}\tr\left(\rho_{\sigma,\epsilon} U_{\sigma,\epsilon}e^{-H_0}\right)\bigg|_{\sigma=\epsilon}\bigg|_{\epsilon=0}\\
    &=Z\frac{\partial}{\partial\epsilon}\frac{\partial}{\partial\sigma}\sum_{n=0}^\infty\frac{(-1)^n}{n!}\tr\left(\rho_{\sigma,\epsilon}U_{\sigma,\epsilon} H_0^n\right)\bigg|_{\sigma=\epsilon}\bigg|_{\epsilon=0}\\
    &=Z\frac{\partial}{\partial\epsilon}\frac{\partial}{\partial\sigma}\sum_{n=0}^\infty\frac{(-1)^n}{2^n n!}h_{a_1b_1}\cdots h_{a_nb_n}\left\langle U_{\sigma,\epsilon}\xi^{a_1}\xi^{b_1}\cdots\xi^{a_n}\xi^{b_n}\right\rangle_{\sigma,\epsilon}
    \bigg|_{\sigma=\epsilon}\bigg|_{\epsilon=0}
    \end{aligned}\label{secondterm}
\end{gather}
Note that we can write $U_{\sigma,\epsilon} = \exp(i h_{\sigma,\epsilon})$, where $h_{\sigma,\epsilon}$ is some Hermitian operator that can be expanded as polynomials of $\xi^a$. Thus expanding the exponential of $U_{\sigma,\epsilon}$ in the above expression, results in even higher-order correlators.
From the higher-order correlators in  \eqref{secondterm} it is evident that non-Gaussian terms will contribute to the second order $S''_0$ term. 

We can derive very general formulas for the entanglement entropy of a system with a single degree of freedom with a generic perturbation. In particular consider the Gaussian state in \eqref{rhog} with the most general perturbation
\begin{equation}
    (\rho_\epsilon)_{q,q'} = \frac 1{Z_\epsilon}\exp\left(-\frac A2(q^2+q'^2) -\frac C2(q-q')^2 + \epsilon f(q,q')\right),
\end{equation}
where $f(x,y)$ is an arbitrary symmetric analytic function of two variables corresponding to the perturbation. For instance the three quartic perturbations considered in Section \ref{quartic} correspond to the functions
\begin{equation}
    f_1(x,y) = x^4+y^4, \quad f_2(x,y) = x^3y+xy^3, \quad \text{and} \quad f_3(x,y) = x^2y^2.
\end{equation}
In Appendix \ref{corrections} we compute $S_0'$ and $S_0''$ for a single degree of freedom with arbitrary perturbations. In particular the second term of \eqref{Sepsilon2}, $(S''_0)_2\!=\!-\tr\left((\rho_\epsilon')^2\rho_\epsilon^{-1}\right)$ which contains the non-Gaussian contributions, is given by
\begin{equation}
    (S''_0)_2 = \frac 1{Z_0}\sqrt{\frac{(2\pi)^3}{\det G}}
    \left[f\!\left(\frac{\partial}{\partial J_1},\frac{\partial}{\partial J_2}\right)-\frac {Z_0'}{Z_0}\right]
    \left[f\!\left(\frac{\partial}{\partial J_2},\frac{\partial}{\partial J_3}\right)-\frac {Z_0'}{Z_0}\right]\exp\left(\frac 12 J^T G^{-1} J\right)\bigg|_{J=0},
\end{equation}
where we have defined
\begin{equation}
    G \equiv 
    \begin{pmatrix}
        0       &   -C     &       C \\
        -C     &   2A+2C    &       -C\\
        C      &   -C     &       0
    \end{pmatrix}, \qquad
    J =
    \begin{pmatrix}
        J_1 \\ J_2 \\ J_3
    \end{pmatrix}.
\end{equation}
Note that $f\!\left(\frac{\partial}{\partial J_1},\frac{\partial}{\partial J_2}\right)$ is a differential operator constructed out of the power expansion of the analytic function $f(x,y)$, where the variables are replaced by differential operators.

\section{Summary, Conclusions, and Outlook}
In this paper we reviewed the main results of \cite{rssee}, where a covariant definition of entropy is proposed for Gaussian theories in terms of the spacetime two-point correlation function. With the aim of generalizing these ideas to the interacting case, we sketched general properties of interacting theories and discussed expectations for the resulting entropy formula. As a first step towards this goal, we considered generic perturbations away from the Gaussian theory. We found that to first order the same formula holds also for non-Gaussian theories with the correlators replaced by their perturbation-corrected versions. 

Naively one would expect higher-order correlators to start contributing as we move away from the Gaussian theory, but this is not the case up to first order of perturbation theory. Only from second order on do we see that higher-order correlators contribute and genuine non-Gaussian effects are observed. The formula in \eqref{secondterm}, expresses these genuine non-Gaussian effects partially in terms of spacetime correlators, and thereby extends the formula in \cite{rssee} to interacting theories (up to second order). In future work, it would be interesting to express this expression entirely and explicitly in terms of spacetime correlators. Such a formulation in terms of spacetime correlators gives us a covariant definition of entropy for interacting theories that is especially useful for the study of  entanglement entropy in general curved spacetimes. 

Furthermore, we also derived closed-form formulas for the corrections of the entanglement entropy in Appendix \ref{corrections}. Path-integral methods such as those in~\cite{Rosenhaus_2014, Rosenhaus_2015, Hertzberg_2012} would be a natural setting to generalize these results to the field theory case.

As a final remark, one might also be able to make an interesting connection between our findings and the horizon molecules in \cite{Molecules}; there may be a relation between horizon molecules being links and the entropy up to first order being fully described by two-point functions.

\acknowledgments
The authors thank Rafael Sorkin, Dorothea Bahns, and Anton van Niekerk for helpful discussions. The authors also thank the Banff International Research Station (BIRS) for hosting the focused research group ``Towards Spacetime Entanglement Entropy for Interacting Theories'' during which the bulk of this research took place. This research was supported in part by Perimeter Institute for Theoretical Physics. Research at Perimeter Institute is supported by the Government of Canada through the Department of Innovation, Science and Economic Development and by the Province of Ontario through the Ministry of Research and Innovation. LH acknowledges support by VILLUM FONDEN via the QMATH center of excellence (grant no.10059). RK is supported by a Chargé des recherches fellowship of the Fonds de la Recherche Scientifique - FNRS (F.R.S.-FNRS), Belgium.  HM acknowledges financial support from ERC Starting Grant No. 678795 TopInSy. YY acknowledges financial support from Imperial College London through an Imperial College Research Fellowship grant, as well as support from the Avadh Bhatia Fellowship at the University of Alberta. MZ acknowledges financial support provided by FCT/Portugal through the IF programme, grant IF/00729/2015.

\appendix
\section{Computation of \texorpdfstring{$p$ and $q$}{} Correlators}

We outline in this appendix the detailed computations of the correlators both for the Gaussian and non-Gaussian theories used in the main text.

\subsection{Correlators for Gaussian Theory}
\label{correlationsg}

We start from the density matrix
\begin{equation}
 \rho_{qq'}=\langle q|\rho|q'\rangle=N_g e^{-A/2(q^2+q'^2)-C/2 (q-q')^2} ,
 \end{equation}
 where $N_g\equiv \sqrt{\frac{A}{\pi}}$ is the normalization constant. With this we obtain
\begin{eqnarray}
    \langle \hat{q}\hat{q}\rangle&=&\tr(\hat{q}^2\rho) %&=&%\int dq\langle q|\hat{q}^2\rho|q\rangle\nonumber\\
    %&=&\int dq\, q^2 \langle q|\rho|q\rangle\nonumber\\
    %&=&
    = N_g \int dq\, q^2 e^{-Aq^2}%\nonumber\\
    %&=&
    = \sqrt{A/\pi}\left(\sqrt{\pi/4A^3}\right)=1/(2A), \\
%\end{eqnarray}
%
%\begin{eqnarray}
    \langle \hat{q}\hat{p}\rangle&=&\tr(\hat{q}\hat{p}\rho)=\int dq_1\, dq_2\,dq_3\,q_{q_1q_2}\,p_{q_2q_3}\,\rho_{q_3q_1}\nonumber\\
    %\int dq\langle q|\hat{q}\hat{p}\rho|q\rangle
    &=& N_g \int dq_1\, dq_2\,dq_3\, q_1\delta(q_1-q_2) \nonumber\\
    &&\times \left(-i\frac{\partial}{\partial q_2}\delta(q_2-q_3)\right)e^{-A/2(q_3^2+q_1^2)-C/2(q_3-q_1)^2}\nonumber\\
    &=&-iN_g\int dq_1\,dq_2\,q_1\delta(q_1-q_2)\frac{\partial}{\partial q_2}e^{-A/2(q_2^2+q_1^2)-C/2(q_2-q_1)^2}\nonumber\\
    %&=&-i N\int dq\, dq'\, q\, \delta(q-q')\frac{\partial}{\partial q'}e^{-A/2 (q'^2+q^2)}\nonumber\\
    &=& iN_g\int dq\, A q^2 e^{-Aq^2}%\nonumber\\
    %&=& i \sqrt{A/\pi} \left(1/2 \sqrt{\pi/A}\right)
    =i/2.
\end{eqnarray}
Since $[\hat{q},\hat{p}]=\hat{q}\hat{p}-\hat{p}\hat{q}=i$,
\begin{eqnarray}
\langle \hat{p}\hat{q}\rangle&=& \langle \hat{q}\hat{p}\rangle- \langle i\rangle=i/2-i=-i/2.
\end{eqnarray}
Finally, for the $\langle \hat{p}\hat{p}\rangle$ correlator,
\begin{eqnarray}
    \langle \hat{p}\hat{p}\rangle&=&\tr(\hat{p}^2\rho)=\int dq_1\, dq_2\,dq_3\, p_{q_1q_2}\,p_{q_2q_3}\,\rho_{q_3q_1}\nonumber\\
    %\int dq\langle q|\hat{p}^2\rho|q\rangle\nonumber\\
    &=& N_g \int dq_1\, dq_2\,dq_3\, \left(-i\frac{\partial}{\partial q_1}\delta(q_1-q_2)\right)  \left(-i\frac{\partial}{\partial q_2}\delta(q_2-q_3)\right)\nonumber\\
    &&\times e^{-A/2(q_3^2+q_1^2)-C/2(q_3-q_1)^2}\nonumber\\
    &=& -N_g\int dq_2\,\left(-(A+C)+A^2q_2^2\right)e^{-Aq_2^2}\nonumber\\
    &=& A+C-A^2\langle \hat{q} \hat{q}\rangle=A/2+C .
   % &=& \int dq'' (A+C-A^2q''^2)\rho=A^2\langle \hat{q}^2\rangle 
    \end{eqnarray}

\subsection{Correlators for non-Gaussian Theory}
\label{corrsng}
Recall that our density matrix is 
\begin{equation}
 \rho_{qq'}=\langle q|\rho|q'\rangle=N e^{-A/2(q^2+q'^2)-C/2 (q-q')^2-\left(\lambda_1 \frac{q^4+q'^4}{2}+\lambda_2(q^3q'+qq'^3)+\lambda_3 q^2q'^2\right)} .
 \end{equation}
With this density matrix we obtain the following correlators (always to first order in $\lambda_i$)
\begin{eqnarray}
    \langle \hat{q}^n\rangle &=& \tr (\hat{q}^n\rho)=N\int dq\, q^n e^{-Aq^2 - (\lambda_1+2\lambda_2+\lambda_3) q^4} \notag \\
    &\simeq& N\int dq\, q^n e^{-Aq^2} \left( 1 - (\lambda_1+2\lambda_2+\lambda_3) q^4 \right) \notag \\
    &=& \frac{N}{N_g} \left( \langle \hat{q}^n\rangle_g -(\lambda_1+2\lambda_2+\lambda_3)\langle \hat{q}^{n+4}\rangle_g \right),
\end{eqnarray}
where the subscript $g$ refers to the correlator in the Gaussian case,
\begin{equation}
\langle \hat{q}^n\rangle_g \equiv N_g \int dq\, q^n e^{-Aq^2}.
\end{equation}
Since, from equation~\eqref{eq:normalization}, $ \frac{N}{N_g} \simeq \left(1+ \frac{3(\lambda_1+2\lambda_2+\lambda_3)}{4A^2} \right)$, we can further write
\begin{eqnarray}
    %\label{eq:corrqn}
    \langle \hat{q}^n\rangle &\simeq& \left(1+ \frac{3(\lambda_1+2\lambda_2+\lambda_3)}{4A^2} \right)
    \left( \langle \hat{q}^n\rangle_g -(\lambda_1+2\lambda_2+\lambda_3)\langle \hat{q}^{n+4}\rangle_g \right) 
    \nonumber \\
    &\simeq& \langle \hat{q}^n\rangle_g + (\lambda_1+2\lambda_2+\lambda_3) \left(
    \frac{3}{4A^2} \langle \hat{q}^n\rangle_g - \langle \hat{q}^{n+4}\rangle_g
    \right) .
\end{eqnarray}
With this we can write the remaining correlators in a more compact form as follows.
\begin{eqnarray}
 \langle \hat{p}^2\rangle &=& \tr(\hat{p}^2\rho)
 = \int dq_1\, dq_2\,dq_3\, p_{q_1q_2}\,p_{q_2q_3}\,\rho_{q_3q_1}\nonumber\\
 &=& N \int dq_1\, dq_2\,dq_3\, \left(-i\frac{\partial}{\partial q_1}\delta(q_1-q_2)\right)  \left(-i\frac{\partial}{\partial q_2}\delta(q_2-q_3)\right)\nonumber\\ 
  && \times e^{-\frac{A}{2}(q_3^2+q_1^2)-\frac{C}{2}(q_3-q_1)^2
  - \lambda_1\frac{q_3^4+q_1^4}{2} - \lambda_2(q_3^3 q_1 + q_3 q_1^3) 
  - \lambda_3 q_3^2 q_1^2 }  \nonumber \\
 &\simeq& \int dq_3\big(A+C-A^2q^2_3
 +6\lambda_1 q_3^2 - 4\lambda_1 A q_3^4 + 6\lambda_2 q_3^2 - 8\lambda_2 A q_3^4
 \nonumber \\
&& \qquad \quad + 2\lambda_3 q_3^2 - 4 \lambda_3 A q_3^4
 \big)\rho_{q_3 q_3}\nonumber\\
 &=&(A+C)-A^2\langle \hat{q}^2\rangle
 + 6\lambda_1 \langle \hat{q}^2\rangle - 4\lambda_1 A \langle \hat{q}^4  \rangle
 + 6\lambda_2 \langle \hat{q}^2 \rangle - 8\lambda_2 A \langle \hat{q}^4 \rangle \nonumber \\
&& + 2\lambda_3 \langle \hat{q}^2 \rangle - 4 \lambda_3 A \langle \hat{q}^4 \rangle \,, \\
     \langle \hat{p}\hat{q}\rangle &=&\int dq_1\, dq_2\,dq_3\, p_{q_1q_2}\,q_{q_2q_3}\,\rho_{q_3q_1}\nonumber\\
     &=& N \int dq_1\, dq_2\, \left(-i\frac{\partial}{\partial q_1}\delta(q_1-q_2)\right) q_2
     \nonumber\\ 
  && \times e^{-\frac{A}{2}(q_2^2+q_1^2)-\frac{C}{2}(q_2-q_1)^2
  - \lambda_1\frac{q_2^4+q_1^4}{2} - \lambda_2(q_2^3 q_1 + q_2 q_1^3) 
  - \lambda_3 q_2^2 q_1^2 }  \nonumber \\
  &\simeq & -i\left(A\langle \hat{q}^2\rangle 
  + 2(\lambda_1+2\lambda_2+\lambda_3) \langle\hat{q}^4\rangle \right)=-i/2 \,,
\end{eqnarray}
and finally we have
\begin{equation}
     \langle \hat{q}\hat{p}\rangle=\overline{\langle \hat{p}\hat{q}\rangle}
     \simeq i\left(A\langle \hat{q}^2\rangle 
     + 2(\lambda_1+2\lambda_2+\lambda_3) \langle\hat{q}^4\rangle \right)=i/2 \,.
\end{equation}

\section{Inverse of M-matrix}\label{appendix:M_inverse}

We are interested in computing the inverse of the matrix \eqref{eq:M_matrix_components}
\begin{equation}
    M = \begin{pmatrix}
            2\beta(1+\mu^2) & 2\beta\mu       & 0           & \cdots    & 0 & 2\beta\mu \\
            2\beta\mu       & 2\beta(1+\mu^2) & 2\beta\mu   &   \cdots  & 0 & 0         \\
                0           & 2\beta\mu       & 2\beta(1+\mu^2) & 2\beta\mu   &   \cdots  & 0  \\
            \vdots          &   \vdots        & \ddots      & \cdots    & \ddots & \vdots\\
            0               &   \cdots        & 0           & 2\beta\mu    & 2\beta(1+\mu^2) & 2\beta\mu \\
            2\beta\mu       &       0         & 0           & \cdots    & 2\beta\mu     & 2\beta(1+\mu^2)
        \end{pmatrix},
\end{equation}
in order to compute correlation functions \eqref{eq:replicaWicks}. There are several ways to do this. One way is to note that the matrix $M$ can be thought of as the matrix elements of a Hamiltonian for a quantum particle on a ring with $n$ sites and nearest-neighbour hopping
\begin{gather}
    \begin{aligned}
        H &= \sum_{j,j'=1}^n M_{jj'}c^\dagger_jc_{j'} \\
            &= \sum_j 2\beta(1+\mu^2) c^\dagger_jc_j - \sum_{\langle ij\rangle}2\beta\mu c^\dagger_jc_{j'},
    \end{aligned}
\end{gather}
where $c_j$ and $c_j^\dagger$ are creation and annihilation operators, respectively.
Since this Hamiltonian has translation symmetry, momentum is a good quantum number. We can therefore find the eigenvalues by a discrete Fourier transform over $\mathbb Z_n$
\begin{equation}
    c_j = \frac 1{\sqrt n}\sum_{k}e^{i kj}c_k,
\end{equation}
giving us the spectrum
\begin{equation}
    \epsilon_k = 2\beta(1+\mu^2) - 4\beta\mu \cos(k), \qquad k = \frac {2\pi}n x,\quad x=0, \dots n-1.
\end{equation}
Since the matrix is diagonal in momentum space, we can invert it
\begin{equation}
    \tilde M^{-1}_{kk'} = \frac 1{\epsilon_k} \delta_{kk'},
\end{equation}
then Fourier transform this inverse matrix back to get
\begin{gather}
    \begin{aligned}
        M^{-1}_{ij} &= \frac 1n\sum_{k,k'} \tilde M^{-1}_{kk'}e^{-ikj}e^{ik'j'}\\
         &=\frac 1n\sum_{x=0}^{n-1}\frac {e^{i\frac {2\pi}n x(j-j')}}{2\beta(1+\mu^2) - 4\beta\mu\cos\left[\frac {2\pi}n x\right]} \,.
    \end{aligned}
\end{gather}
Alternatively, we can use something similar to a Cholesky decomposition and write $M = Q^TQ$ where
\begin{equation}
    Q = \sqrt{2\beta}\begin{pmatrix}
            1 & -\mu &           &          &   &\\
              &   1  & -\mu      &          &   &\\
              &      &   \ddots  &  \ddots  &   &\\
            -\mu  &      &           &      & 1 & -\mu
        \end{pmatrix}.
\end{equation}
We can then write the inverse as $M^{-1} = Q^{-1}\left(Q^{-1}\right)^T$. The inverse of $Q$ is given by
\begin{equation}
    Q^{-1}_{ij} = \frac{\mu^{(j-i) \!\mod n}}{(1-\mu^n)\sqrt{\beta}}.
\end{equation}
From this we get another expression for the inverse of $M$
\begin{equation}
    M^{-1}_{ij}=\sum_{k=1}^n\frac{\mu^{(k-i)\Mod{n} + (k-j)\Mod{n}}}{2(1-\mu^n)^2\beta}.
\end{equation}
This matrix is dense, but for our purposes we only need the diagonal
\begin{equation}
    M^{-1}_{ii}=\sum_{k=0}^{n-1}\frac{\mu^{2k}}{2(1-\mu^n)^2\beta}=\frac{(\mu^{2n}-1)}{2(1-\mu^n)^2\beta(\mu^2-1)},
\end{equation}
and the next-to-diagonal elements
\begin{equation}
    M^{-1}_{ii+1}=M^{-1}_{ii-1}=\frac{\sum_{k=0}^{n-2}\left(\mu^{2k+1}+\mu^{n-1}\right)}{2(1-\mu^n)^2\beta}=\frac{(\mu^n+\mu^2)}{2\mu\beta(\mu^2-1)(\mu^n-1)}.
\end{equation}
Here we have used that $M_{ii}=M_{11}$ and $M_{ii+1} = M_{12}$ for any $i$.
The determinant is given by
\begin{equation}
    \det M=2^n(1-(-\mu)^n)^2\beta^n.
\end{equation}

\section{Peierls Bracket for Interacting Theories and non c-number Commutators}\label{appendix:c_numberStuff}

We will here see a simple example of how interacting theories gain non c-number Peierls brackets and commutators (of Heisenberg operators).\footnote{See also reference \cite{YangFeld1}.} Consider the equations of motion for a $\phi^{p+1}$ theory
\begin{equation}
    \left(\Box + m^2\right)\phi(x) = \lambda\phi^p(x).
\end{equation}
We can perturbatively solve these equations with the following ansatz
\begin{equation}
    \phi(x) = \sum_n\lambda^n\phi_n(x),
\end{equation}
where each term satisfies the following differential equations
\begin{equation}
    \left(\Box + m^2\right)\phi_0 = 0, \qquad \left(\Box + m^2\right)\phi_n = \sum_{n_1+\cdots+ n_p+1=n}\phi_{n_1}\cdots\phi_{n_p}.
\end{equation}
The sum should be understood as the sum of all possible choices of $n_1, n_2, \dots n_p$ such that $n_1+\cdots+ n_p+1=n$.
We can express the solution in terms of the solution of the unperturbed theory to, say, first order as
\begin{equation}\label{eq:perturbativesolutionPhiP}
    \phi(x) = \phi_0(x) + \lambda \int dy\, G_R(x,y)\phi_0^p(y) + \mathcal O(\lambda^2),
\end{equation}
where by $G_R$ and $G_A$ we will denote the retarded and advanced Green functions.
The Peierls bracket of the unperturbed theory is given by \cite{peierls}
\begin{equation}
    \left\{\phi_0(x), \phi_0(y)\right\} = \Delta(x,y),
\end{equation}
where $\Delta = G_R - G_A$. Quantizing this theory, we will find c-number commutators of the Heisenberg opeators. Using this Peierls bracket together with \eqref{eq:perturbativesolutionPhiP}, we  find the Peierls bracket of the interacting field to be of the form
\begin{equation}
    \left\{\phi(x), \phi(y)\right\} = \Delta(x,y) + \lambda\times(\text{non c-number terms}) + \mathcal O(\lambda^2).
\end{equation}
Here by ``non c-number'' terms we mean those containing polynomials of $\phi_0$. Quantizing the interacting theory requires some care as the normal ordering of the non c-number terms is ambiguous and must be chosen in a way that avoids any anomalies. The above illustrates the connection between adding interactions in the Lagrangian and the appearance of non c-number corrections in the interacting field commutator.

Generally, theories with c-number commutators are (generalized) free theories \cite{Greenberg} and those that have non c-number terms correspond to interacting theories. Actually, if any truncated (connected) Wightman function of order $2n$ ($n>1$) is zero then the theory turns out to a be generalized free theory \cite{Baumann}. Thus interacting theories have non c-number commutators and all their even truncated Wightman functions are non-zero.

\section{\texorpdfstring{$S_0'$ and $S_0''$}{S0 and S0''} for a Single Degree of Freedom and Arbitrary Perturbations}
\label{corrections}
Consider a general perturbation \begin{equation}\label{eq:GeneralPerturbedStateWithf}
    (\rho_\epsilon)_{q,q'} = \frac 1{Z_\epsilon}\exp\left(-\frac A2(q^2+q'^2) -\frac C2(q-q')^2 + \epsilon f(q,q')\right),
\end{equation}
for any analytic symmetric function of two variables $f(x,y)$. For the derivation we also need the Gaussian part in operator form, which is given by
\begin{equation}
    \hat\rho_0 = \frac 1{\widetilde Z_0}\exp\left(K\left[\alpha \hat q^2 + \beta \hat p^2\right]\right),
\end{equation}
where
\begin{equation}\label{eq:alphaBetaCorrelatorDef}
    \beta=\langle\hat q\hat q\rangle = \frac 1{2A}, \qquad \alpha=\langle\hat p\hat p\rangle = \frac A2 +C,
\end{equation}
and
\begin{equation}
    K = -\frac 1{2\sigma}\log\left(\frac{\sigma+\frac 12}{\sigma-\frac 12}\right), \qquad \widetilde Z_0 = \sqrt{\left(\sigma-\frac 12\right)\left(\sigma+\frac 12\right)}.
\end{equation}
Note that $\sigma^2=\det R/\det\Delta$, or more specifically given by \eqref{sigmaca}. Let us consider the perturbative expansion of the entropy  
\begin{equation}
    S_\epsilon = S_0 + \epsilon S'_0 + \epsilon^2 S''_0 + \mathcal{O}(\epsilon^3).
\end{equation}
As discussed in the main text, the first two terms are given by
	\begin{equation}
		S_0'=-\mathrm{tr}\left(\rho_0'\log\rho_0\right) \qquad\text{and}\qquad S_0''=-\mathrm{tr}\left(\rho_0''\log\rho_0+\rho_0'\rho_0^{-1}\rho_0'\right).
	\end{equation}
In the following we will evaluate these expressions for the general perturbed state \eqref{eq:GeneralPerturbedStateWithf} in terms of the perturbation function $f(x,y)$.

\subsection{First Order, \texorpdfstring{$S_0'$}{S0'}}
The first order correction is

\begin{equation}\label{eq:S0PrimeFirstOrderSection}
    S_0'=-K\,\mathrm{tr}\left(\rho_0'\left[\alpha \hat q^2 + \beta \hat p^2\right]\right) - \log\widetilde Z_0.
\end{equation}
One can easily compute the derivative of the density matrix, which is given by
\begin{equation}\label{eq:rhoPrimeZero}
    (\rho'_0)_{q,q'} = \left[f(q,q')-\frac{Z'_0}{Z_0}\right](\rho_0)_{q,q'},
\end{equation}
where
\begin{gather}
    \begin{aligned}
    Z^{(n)}_0 &= \frac{d^n}{d\epsilon^n}\tr\left(\rho_{\epsilon}\right)\bigg|_{\epsilon=0}=
    \int dq f^n(q,q) e^{-A q^2},\\
            &=\sqrt{\frac{\pi}A} f^n\left(\frac{\partial}{\partial J},\frac{\partial}{\partial J}\right)e^{J^2/4A}\bigg|_{J=0}.
    \end{aligned}
\end{gather}
This expression is derived by first computing the Gaussian integral with a linear current term $g[J] \equiv \int dq\, e^{-A q^2+Jq} = \sqrt{\frac \pi A}e^{J^2/4A}$. Note that $\frac{\partial^n}{\partial J^n}g[J]|_{J=0} = \int dq \,q^n e^{-A q^2}$. Since we assumed that $f(x,y)$ is analytic, it implies that $f(q,q)$ is a polynomial or a power-series and thus $f^n\left(\frac{\partial}{\partial J},\frac{\partial}{\partial J}\right)g[J]|_{J=0} = \int dq \,f^n(q,q) e^{-A q^2}$. We will use this trick several times in the following.

All we have to do now is to insert \eqref{eq:rhoPrimeZero} into \eqref{eq:S0PrimeFirstOrderSection} and evaluate the trace in the position basis. Some of the terms are just two-point functions \eqref{eq:alphaBetaCorrelatorDef}, while others can be evaluated using the above mentioned trick (using integration by parts). The result is
\begin{equation}
    S'_0 = K\,\hat{\mathcal D_1}e^{J^2/(4A)}\bigg|_{J=0},
\end{equation}
where we have defined the differential operator

\begin{gather}
    \begin{aligned}
    \hat{\mathcal D}_1 &\equiv -(A+C)f\left(\frac{\partial}{\partial J},\frac{\partial}{\partial J}\right)\frac{\partial^2}{\partial J^2} + \frac 1{2A}\left[\frac{\partial^2}{\partial y^2}f(x,y)\right]\bigg|_{x=y=\frac{\partial}{\partial J}}
    \\
    &-\left[\frac{\partial}{\partial y}f(x,y)\right]\bigg|_{x=y=\frac{\partial}{\partial J}}\frac{\partial}{\partial J}
    - \left[\sigma-\frac 34 - \frac{\log \widetilde Z_0}K (Z_0-1) \right]f\left(\frac{\partial}{\partial J},\frac{\partial}{\partial J}\right).
    \end{aligned}
\end{gather}

\subsection{Second Order, \texorpdfstring{$S_0''$}{S0''}}
The second order term can be split into two contributions
\begin{equation}
    S''_0 = (S''_0)_1 + (S''_0)_2.
\end{equation}
The first contribution is given by 
\begin{equation}
    (S''_0)_1 = -\mathrm{tr}\left(\rho_0''\log\rho_0\right)
\end{equation}
and as argued earlier, this term is only sensitive to the Gaussian part of our state. The second derivative of the density matrix is given by
\begin{equation}
    (\rho''_0)_{q,q'} = \left[\left(f(q,q')-2\frac{Z'_0}{Z_0}\right)f(q,q')+ \left(-\frac{Z''_0}{Z_0}+2\left(\frac{Z'_0}{Z_0}\right)^2\right)f(q,q')\right](\rho_0)_{q,q'}.
\end{equation}
The calculation then proceeds in exactly the same way as for the first order term. The `Gaussian' contribution to the second order term is
\begin{equation}
    (S''_0)_1 = -\sqrt{\frac \pi A}K\,\hat{\mathcal D_2}e^{J^2/(4A)}\bigg|_{J=0},
\end{equation}
where

\begin{gather}
    \begin{aligned}
    \hat{\mathcal D_2} &\equiv \left(\frac A2+C\right)\left[f^2\left(\dJ,\dJ\right) - 2 f\left(\dJ, \dJ\right)\frac{Z_0'}{Z_0}\right]\dJn 2
    - \frac 1{2A}\Bigg[
        2\diff{y}\left\{\left(f(x,y)-\frac{Z'_0}{Z_0}\right)\diff{y}f(x,y)\right\}
        \\&- 4A \left(f(x,y)-\frac{Z_0'}{Z_0}\diff{y}f(x,y)\right)
        +\left(A^2 x^2-A-C\right)\left\{f^2(x,y) - 2f(x,y)\frac{Z_0'}{Z_0}\right\}
    \Bigg]\Bigg|_{x=y=\dJ}\\
    &
    -\frac{\log\widetilde Z_0}K\left[\left(1+\sqrt{\frac A\pi}\right)f^2\left(\dJ,\dJ\right)-2f\left(\dJ,\dJ\right)\frac{Z_0'}{Z_0}\right] + \frac{A+2C}{2K\sqrt{A\pi}}f^2\left(\dJ,\dJ\right)
    \end{aligned}
\end{gather}

The second part of the second order term contains all the non-Gaussian contributions and is given by
\begin{equation}
    (S_0'')_2=-\mathrm{tr}\left(\rho_0'\rho_0^{-1}\rho_0'\right).
\end{equation}
This term can be evaluated directly as integrals in the $q$-basis. The same Gaussian integral trick we used earlier can also be used here, although now on a three-dimensional Gaussian integral. The result is
\begin{equation}
    (S''_0)_2 = \frac 1{Z_0}\sqrt{\frac{(2\pi)^3}{\det G}}
    \left[f\left(\frac{\partial}{\partial J_1},\frac{\partial}{\partial J_2}\right)-\frac {Z_0'}{Z_0}\right]
    \left[f\left(\frac{\partial}{\partial J_2},\frac{\partial}{\partial J_3}\right)-\frac {Z_0'}{Z_0}\right]\exp\left(\frac 12 J^T G^{-1} J\right)\bigg|_{J=0},
\end{equation}
where we have defined
\begin{equation}
    G \equiv 
    \begin{pmatrix}
        0       &   -C     &       C \\
        -C     &   2A+2C    &       -C\\
        C      &   -C     &       0
    \end{pmatrix}, \qquad
    J =
    \begin{pmatrix}
        J_1 \\ J_2 \\ J_3
    \end{pmatrix}.
\end{equation}

\bibliography{references}{}
\bibliographystyle{JHEP}

\end{document}